\documentclass[aps,prl,preprint,tightenlines,superscriptaddress,showpacs,byrevtex]{revtex4}
\usepackage{graphicx} 
\usepackage{epsfig}
\usepackage{dcolumn}  
\usepackage{color}
\usepackage{graphicx}
\usepackage{rotating}

\definecolor{red}{rgb}{1,0,0}
\definecolor{green}{rgb}{0,0.7,0}
\definecolor{violet}{rgb}{0.4,0.0,0.55}
\definecolor{blue}{rgb}{0,0,0.7}
\definecolor{magenta}{rgb}{0.8,0.0,0.5}
\definecolor{ltblue}{rgb}{0.000,0.750,0.750}

\newcommand{\beq}{\begin{eqnarray}}
\newcommand{\eeq}{\end{eqnarray}}
\newcommand{\bem}{\begin{pmatrix}}
\newcommand{\eem}{\end{pmatrix}}

\newcommand{\BB}{\ensuremath{B\bar{B}}}

\newcommand{\Bbaro}{\ensuremath{\bar{B}{}^0}}

\newcommand{\de}{\ensuremath{\Delta E}}
\newcommand{\mb}{\ensuremath{M_{\mbox{\scriptsize bc}}}}
\newcommand{\etap}{\ensuremath{\eta^{\prime}}}

\newcommand{\BF}{\ensuremath{{\mathcal B}}}

\newcommand{\babar}{{\sc BaBar}}
\newcommand{\LK}{\ensuremath{{\cal L}}}

\newcommand{\FD}{\ensuremath{{\cal F}}}
\newcommand{\LR}{\ensuremath{{\cal R_L}}}

\newcommand{\cost}{\ensuremath{\cos\theta_T}}
\newcommand{\cosb}{\ensuremath{\cos\theta_B}}
\newcommand{\sperp}{\ensuremath{S_\perp}}
\newcommand{\ebeam}{\ensuremath{E_{\mbox{\scriptsize beam}}}}
\newcommand{\cs}{\ensuremath{/c^2}}

\newcommand{\eff}{\ensuremath{\epsilon}}

\newcommand{\etagg}{\ensuremath{\eta_{\gamma\gamma}}}

\newcommand{\etaorp}{\ensuremath{\eta^{(\prime)}}}

\newcommand{\bteprhoc}{\ensuremath{B^{(0,+)} \to \etap \rho^{(0,+)}}}
\newcommand{\btepkstarc}{\ensuremath{B^{(0,+)} \to \etap K^{*(0,+)}}}
\newcommand{\btepphic}{\ensuremath{B^0 \to \etap \phi}}
\newcommand{\btepomegac}{\ensuremath{B^0 \to \etap \omega}}
\newcommand{\btepetaorpc}{\ensuremath{B^0 \to \etap \etaorp}}

\newcommand{\btephh}{\ensuremath{B \to \etap h^{(*)}}}
\newcommand{\bteprho}{\ensuremath{B \to \etap \rho}}
\newcommand{\bteprhoo}{\ensuremath{B^0 \to \etap \rho^0}}
\newcommand{\bteprhop}{\ensuremath{B^+ \to \etap \rho^+}}
\newcommand{\btepkstar}{\ensuremath{B \to \etap K^*}}
\newcommand{\btepkstaro}{\ensuremath{B^0 \to \etap K^{*0}}}
\newcommand{\btepkstarp}{\ensuremath{B^+ \to \etap K^{*+}}}
\newcommand{\btepphi}{\ensuremath{B^0 \to \etap \phi}}
\newcommand{\btepomega}{\ensuremath{B^0 \to \etap \omega}}
\newcommand{\btepeta}{\ensuremath{B^0 \to \etap \eta}}
\newcommand{\btepetap}{\ensuremath{B^0 \to \etap \etap}}
\newcommand{\btepetaorp}{\ensuremath{B^0 \to \etap \etaorp}}
\newcommand{\btepk}{\ensuremath{B \to \etap K}}
\newcommand{\bteppi}{\ensuremath{B \to \etap \pi}}
\newcommand{\btepkp}{\ensuremath{B^+ \to \etap K^+}}

\newcommand{\epp}{\ensuremath{\etap \to \eta \pi^+ \pi^-}}
\newcommand{\erg}{\ensuremath{\etap \to \rho^0 \gamma}}

\newcommand{\nepp}{\ensuremath{\eta \pi \pi}}
\newcommand{\nerg}{\ensuremath{\rho \gamma}}

\begin{document}

\vspace*{-3\baselineskip}
\resizebox{!}{3cm}{\includegraphics{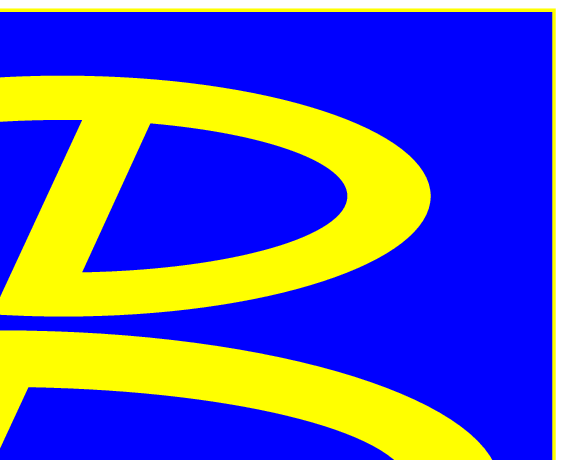}}

\preprint{\vbox{ \hbox{KEK-PREPRINT 2006-70}
                 \hbox{BELLE-PREPRINT 2006-7}
                 \hbox{hep-ex 0701046}
}}

\title{ \quad\\[0.5cm]  Search for $B$ decays into $\etap \rho$, $\etap K^*$,
$\etap\phi$, $\etap\omega$ and $\etap\etaorp$}


\affiliation{University of Cincinnati, Cincinnati, Ohio 45221}
\affiliation{The Graduate University for Advanced Studies, Hayama}
\affiliation{Hanyang University, Seoul}
\affiliation{University of Hawaii, Honolulu, Hawaii 96822}
\affiliation{High Energy Accelerator Research Organization (KEK), Tsukuba}
\affiliation{Hiroshima Institute of Technology, Hiroshima}
\affiliation{Institute of High Energy Physics, Vienna}
\affiliation{Institute of High Energy Physics, Protvino}
\affiliation{Institute for Theoretical and Experimental Physics, Moscow}
\affiliation{J. Stefan Institute, Ljubljana}
\affiliation{Kanagawa University, Yokohama}
\affiliation{Korea University, Seoul}
\affiliation{Kyungpook National University, Taegu}
\affiliation{Swiss Federal Institute of Technology of Lausanne, EPFL, Lausanne}
\affiliation{University of Ljubljana, Ljubljana}
\affiliation{University of Maribor, Maribor}
\affiliation{University of Melbourne, Victoria}
\affiliation{Nagoya University, Nagoya}
\affiliation{Nara Women's University, Nara}
\affiliation{National Central University, Chung-li}
\affiliation{National United University, Miao Li}
\affiliation{Department of Physics, National Taiwan University, Taipei}
\affiliation{H. Niewodniczanski Institute of Nuclear Physics, Krakow}
\affiliation{Nippon Dental University, Niigata}
\affiliation{Niigata University, Niigata}
\affiliation{University of Nova Gorica, Nova Gorica}
\affiliation{Osaka City University, Osaka}
\affiliation{Osaka University, Osaka}
\affiliation{Panjab University, Chandigarh}
\affiliation{Peking University, Beijing}
\affiliation{Princeton University, Princeton, New Jersey 08544}
\affiliation{RIKEN BNL Research Center, Upton, New York 11973}
\affiliation{Seoul National University, Seoul}
\affiliation{Shinshu University, Nagano}
\affiliation{Sungkyunkwan University, Suwon}
\affiliation{University of Sydney, Sydney New South Wales}
\affiliation{Tata Institute of Fundamental Research, Bombay}
\affiliation{Toho University, Funabashi}
\affiliation{Tohoku Gakuin University, Tagajo}
\affiliation{Tohoku University, Sendai}
\affiliation{Department of Physics, University of Tokyo, Tokyo}
\affiliation{Tokyo Institute of Technology, Tokyo}
\affiliation{Tokyo Metropolitan University, Tokyo}
\affiliation{Tokyo University of Agriculture and Technology, Tokyo}
\affiliation{Virginia Polytechnic Institute and State University, Blacksburg, Virginia 24061}
\affiliation{Yonsei University, Seoul}
  \author{J.~Sch\"umann}\affiliation{National United University, Miao Li} 
  \author{C.~H.~Wang}\affiliation{National United University, Miao Li} 
  \author{K.~Abe}\affiliation{High Energy Accelerator Research Organization (KEK), Tsukuba} 
  \author{H.~Aihara}\affiliation{Department of Physics, University of Tokyo, Tokyo} 
  \author{D.~Anipko}\affiliation{Budker Institute of Nuclear Physics, Novosibirsk} 
  \author{K.~Arinstein}\affiliation{Budker Institute of Nuclear Physics, Novosibirsk} 
  \author{V.~Aulchenko}\affiliation{Budker Institute of Nuclear Physics, Novosibirsk} 
  \author{T.~Aushev}\affiliation{Swiss Federal Institute of Technology of Lausanne, EPFL, Lausanne}\affiliation{Institute for Theoretical and Experimental Physics, Moscow} 
  \author{A.~M.~Bakich}\affiliation{University of Sydney, Sydney New South Wales} 
  \author{E.~Barberio}\affiliation{University of Melbourne, Victoria} 
% ...
\collaboration{The Belle Collaboration}
\noaffiliation

\begin{abstract}
We report on a search for the exclusive two-body charmless hadronic $B$ meson
decays \bteprho , \btepkstar , \btepphi , \btepomega{} and \btepetaorp .
The results are obtained from a data sample containing 535
$\times 10^6$ \BB{} pairs that were collected 
at the $\Upsilon(4S)$ resonance
with the Belle detector at the KEKB asymmetric-energy $e^+ e^-$
collider.
We find no significant signals and report upper limits 
in the range $(0.5$--$6.5)\times 10^{-6}$
for all of the above decays.
\end{abstract}


\maketitle

\tighten

\section{Introduction}
Information on the two-body charmless hadronic $B$ meson decays with an \etap{}
meson in the final state (\btephh ) 
is incomplete at present. While the decay \btepk{} is
observed with a large branching fraction, so far no other \btephh{} decay mode
has been observed with greater than $5\sigma$ significance. The first 
evidence of \bteppi{} has recently been reported
~\cite{Aubert:2005bq,Schuemann} and \babar{} found evidence for \btepkstar{}
with larger than $4 \sigma$ significance~\cite{Aubert:2006as}, 
and thus additional observations are expected in the near future.
The study of these decay modes can improve the understanding
of the flavor-singlet penguin amplitude with intermediate
$t$, $c$ and $u$ quarks~\cite{Chiang:2004nm}. Furthermore, these studies 
increase our confidence
in the reliability of a variety of other predictions, e.g, for the $CP$ 
violating parameter $\phi_3$ ($\gamma$), and are necessary to extract
theory parameters such as the scalar penguin
operator~\cite{Chiang:2004nm,Beneke:2003zv}. 
Presently, theoretical predictions for the branching
fractions of these decay modes cover the range
(0.0001--7.6)$\times 10^{-6}$~\cite{Chiang:2004nm,Beneke:2003zv,Fu:2003fy}.
The most stringent upper limits for presently unobserved decays
were reported by \babar~\cite{Aubert:2005bq,Aubert:2006qd,Aubert:2006as}.

\section{Data set and apparatus}
The study performed here includes the decays \bteprhoc, \btepkstarc, \btepphic,
\btepomegac{} and \btepetaorpc{} and 
is based on a data sample that
contains 535 $\times 10^6$ \BB{} pairs, 
collected  with the Belle detector at the KEKB asymmetric energy
$e^+e^-$ (3.5~GeV on 8~GeV) collider~\cite{KEKB}.
Throughout this paper, 
the inclusion of the charge conjugate decay is implied
unless stated otherwise.

KEKB operates at the $\Upsilon(4S)$ resonance 
($\sqrt{s}=10.58$~GeV) with a peak luminosity that exceeds
$1.7\times 10^{34}~{\rm cm}^{-2}{\rm s}^{-1}$.
The Belle detector is a large-solid-angle magnetic
spectrometer that
consists of a silicon vertex detector (SVD),
a 50-layer central drift chamber (CDC), an array of
aerogel threshold \v{C}herenkov counters (ACC), 
a barrel-like arrangement of time-of-flight
scintillation counters (TOF), and an electromagnetic calorimeter
comprised of CsI(Tl) crystals located inside 
a superconducting solenoid coil that provides a 1.5~T
magnetic field.  An iron flux-return located outside of
the coil is instrumented to detect $K_L^0$ mesons and to identify
muons.  The detector
is described in detail elsewhere~\cite{Belle}.
Two inner detector configurations were used. A 2.0 cm beampipe
and a 3-layer SVD were used for the first data sample
of 152 $\times 10^6$ \BB{} pairs (Set $I$), while a 1.5 cm beampipe, a 4-layer
SVD and a small-cell inner drift chamber were used to record  
the remaining 383 $\times 10^6$ \BB{} pairs (Set $II$)\cite{Ushiroda}.  

\section{Event selection and reconstruction}
For what follows, unless stated otherwise, 
all variables are defined in the center-of-mass
frame with the $z$ axis anti-parallel to the positron direction. 

Charged hadrons are identified 
by combining information from the CDC ($dE/dx$),
ACC and TOF systems. Both kaons and pions are selected with an average 
efficiency of 86\% and are misidentified as pions or kaons, respectively,
in 4\% of the cases.

The \etap{} mesons are reconstructed in the decays
\epp{} (with $\eta \to \gamma \gamma)$ and \erg, except for
the decays \btepeta, 
\btepetap{} and \btepomega, which use only the \epp{} channel.
We define the \etap{} ($h^{(*)}$) side as all particles involved in the decay of
the \etap{} ($h^{(*)}$) from the decay \btephh .
The $\eta$, $\rho^0$ and \etap{} candidates on the 
\etap{} side are reconstructed
using the mass windows given in Table~\ref{tab:etapcuts}. Mass windows used to
reconstruct the $h^{(*)}$ are given in Table~\ref{tab:cuts}. 
In addition, we require
the following: photons originating from $\pi^0$ and $\eta$ decays are required
to have energies of at least 100 MeV, photons from the \etap{} in \erg{} have to
be above 200 MeV in the laboratory frame.
The transverse momenta of the $\pi^\pm$ for $\etap \to \etagg\pi^+\pi^-$ 
($\etap \to \rho^0_{\pi^+ \pi^-} \gamma$)
candidates have to be greater than 100 MeV/$c$ (200 MeV/$c$).
An additional requirement on the cosine of the $\rho^0$ helicity angle 
in \erg{} of 
$|\cos\theta_h| < 0.85$ is applied, 
where $\theta_h$ is the angle between the momenta 
of one of the daughter pions of the
$\rho^0$ and the $\etap$ in the $\rho^0$ rest frame.
The vertex of the
$K_S^0\to\pi^+\pi^-$ has to be displaced from the interaction point (IP)
and the $K_S^0$ momentum direction must be consistent with its flight
direction as 
indicated in Table~\ref{tab:KS}~\cite{bib:ks}. 
\begin{table}[htb!]
\caption{Invariant mass windows used to select 
intermediate states on the \etap{} side.
$\sigma$ denotes a standard deviation of the reconstructed mass distribution.}
\label{tab:etapcuts}
\begin{tabular}
{@{\hspace{0.5cm}}l@{\hspace{0.5cm}}@{\hspace{0.5cm}}c@{\hspace{0.5cm}}
@{\hspace{0.5cm}}c@{\hspace{0.5cm}}}
\hline \hline
Mode	& \multicolumn{2}{c}{Mass window}	\\
	&	 (MeV\cs )	& 	in units of $\sigma$ 	\\
\hline
$\rho^0\to\pi^+\pi^-$  	&	[550,870] & ---		\\
$\eta\to\gamma\gamma$  	&	[500,570] & $+2.5/-3.3$  \\
$\epp$ 			&	[950,965] & $\pm 2.5 $	\\
$\erg$ 			&	[941,970] & $\pm 2.5 $	\\
\hline \hline
\end{tabular}
\end{table}
\begin{table}[htb!]
\caption{Invariant mass windows used to select 
intermediate states on the $h^{(*)}$ side.
$\sigma$ denotes a standard deviation of the reconstructed mass distribution.}
\label{tab:cuts}
\begin{tabular}
{@{\hspace{0.5cm}}l@{\hspace{0.5cm}}@{\hspace{0.5cm}}c@{\hspace{0.5cm}}
@{\hspace{0.5cm}}c@{\hspace{0.5cm}}}
\hline \hline
Mode	& \multicolumn{2}{c}{Mass window}	\\
	&	 (MeV\cs )	& 	in units of $\sigma$ 	\\
\hline
$\pi^0\to\gamma\gamma$	&	[118,150] & $\pm 2.5 $  \\
$K_S^0\to\pi^+\pi^-$	& 	[485,510] & $\pm 3 $ 	\\
$\rho^+\to\pi^+\pi^0$  	&	[620,920] & ---		\\
$\rho^0\to\pi^+\pi^-$  	&	[620,920] & ---		\\
$K^{*0}\to K^+\pi^-$	&	[820,965] & ---		\\
$K^{*+}\to K_S^0\pi^+$	&	[820,965] & ---		\\
$K^{*+}\to K^+\pi^0$	&	[820,965] & ---		\\
$\phi\to K^+ K^-$	&	[1010,1030] & $\pm 3 $	\\
\hline
$\eta\to\gamma\gamma$  	&	[510,575] & $\pm 2.5 $ 	\\
$\epp$ 			&	[950,965] & $\pm 2.5 $ 	\\
$\omega\to\pi^+\pi^-\pi^0$ &	[750,810] & $\pm 2.5 $ 	\\
\hline \hline
\end{tabular}
\end{table}
\begin{table*}[htb!]
\caption{Selection criteria for the distance of closest approach of one of the 
$K_S^0$ daughter pions to the IP ($dr$) in the $x$-$y$ plane, azimuthal angle
between the momentum vector and the flight direction of the $K_S^0$ candidate
inferred from the production and decay vertexes ($d\phi$),
distance of closest approach between the two daughter tracks ($z$-dist.) and 
the flight length of the $K_S^0$ candidate in the $x$-$y$ plane ($fl$).}
\label{tab:KS}
\begin{tabular}
{@{\hspace{0.5cm}}l@{\hspace{0.5cm}}@{\hspace{0.5cm}}c@{\hspace{0.5cm}}
@{\hspace{0.5cm}}c@{\hspace{0.5cm}}@{\hspace{0.5cm}}c@{\hspace{0.5cm}}
@{\hspace{0.5cm}}c@{\hspace{0.5cm}}}
\hline \hline
Momentum (GeV/$c$)	&	$dr$ (cm)	&	$d\phi$(rad)	&	$z$-dist. (cm)	&	$fl$ (cm)	\\
\hline
\hspace{0.1cm} $<0.5$ &	$>0.05$		&	$<0.3$		&	$<0.8$		&	---	\\
$0.5-1.5$	&	$>0.03$		&	$<0.1$		&	$<1.8$		&	$>0.08$	\\
\hspace{0.1cm} $>1.5$ &	$>0.02$		&	$<0.03$		&	$<2.4$		&	$>0.22$	\\
\hline \hline
\end{tabular}
\end{table*}

$B$ meson candidates are formed by combining an $\etap$ meson with 
one of the hadrons listed in Table~\ref{tab:cuts} excluding $\pi^0$'s and
$K_S^0$'s. 
$B$ candidates are identified
using two kinematic variables: the energy difference, $\de = E_B-\ebeam$, 
and the beam-energy constrained
mass, $\mb =\sqrt{\ebeam^2/c^4 - (P_B/c)^2}$, 
where \ebeam{} is the beam energy and $E_B$
($P_B$) is the reconstructed energy (momentum) of the $B$ candidate. 
Signal events peak at $\de = 0$ GeV and $\mb = M_B$, where $M_B$ is the
$B$ meson mass, with resolutions
around $15$ MeV and $3$ MeV for \de{} and \mb{} respectively. An $\eta$ mass
constraint fit is applied in the \epp{} subdecay in order 
to improve the \de{} resolution.
Here the two photons from $\eta \to \gamma \gamma$ are constrained to have the
nominal $\eta$ mass given by the Particle Data Group (PDG)~\cite{bib:PDG04}.
Events satisfying the requirements $\mb > 5.22$
GeV\cs{} and $|\de| <0.25$ GeV are selected for further analysis. After all
selections are applied,  depending on the decay mode
between 3\% and 20\% of the events have multiple $B$ candidates in one event. 
A $\chi^2$ variable is calculated to select the best candidate of such
events. We select the $B$ with the smallest 
$\chi^2 = \chi^2_{\text{vtx}} + \chi^2(M_{\etap}) + \chi^2(M_{h^{(*)}})$, 
where $\chi^2_{\text{vtx}}$ 
is an estimator of the vertex quality for all charged particles not from the
$K_S^0$ and $\chi^2(M_{X}) =[(M_{X}-m_{X})/\sigma_{X}]^2$,
where $M_{X}$ ($m_X$) is the reconstructed (nominal) mass of the particle 
candidate $X$ ($=$ \etap{} or $h^{(*)}$) and $\sigma_{X}$
is the standard deviation of the reconstructed $X$ mass distribution
as obtained from fits to MC distributions.

\section{Background suppression}
The dominant background for this analysis is continuum 
$e^+e^- \to q\bar q$ ($q =u,$ $d,$ $s,$ $c$). Other background sources are
charmless $B$ decays such as \btepk{} and $b\to c$ decays. The background
is 90\% continuum with the remaining 10\%
nearly evenly split between the other two contributions.
 
Several event shape variables are used to distinguish
the spherical \BB{} topology from 
the jet-like $e^+e^- \to q\bar q$ continuum background. 
The thrust angle $\theta_T$ is defined
as the angle between the \etap{} momentum direction and 
the thrust axis formed by all particles not belonging to the reconstructed $B$
meson.
Continuum events tend to peak near $|\cost| = 1$,
while \BB{} events have a uniform distribution.
The requirement $|\cost|<0.9$ is applied prior
to all other event topology selections resulting in a signal efficiency
(background reduction) of 90\% (56\%).

Additional continuum background suppression is obtained by using
modified Fox-Wolfram moments~\cite{SFW} and $|\cosb|$, where
$\theta_B$ is the angle between the flight 
direction of the reconstructed $B$ candidate and the beam axis. A 
Fisher discriminant (\FD)~\cite{fisher:1936} is formed from a linear 
combination of $|\cost|$, $\sperp$~\cite{Ammar:1993sh} 
and five modified Fox-Wolfram moments.
$\sperp$ is the ratio of
the scalar sum of the transverse momenta of all tracks outside a 
$45^{\circ}$ cone around the $\etap$ direction 
to the scalar sum of their total momenta.
The Fisher discriminant is then combined with the $B$ flight direction
information to form an event topology likelihood function
$\LK_S$ ($\LK_{q\bar{q}}$), 
where the subscript $S$ ($q\bar{q}$) represents signal (continuum background).
The signal over continuum background ratio varies over the range of the
quality parameter $r$ of the $B$ flavor tagging of the accompanying $B$ meson.
We use the 
standard Belle $B$ tagging algorithm~\cite{TaggingNIM}, 
which gives the $B$ flavor
and the tagging quality $r$ ranging from zero for no flavor to unity
for unambiguous flavor assignment. The data is divided into three
$r$ regions and the likelihood ratio $\LR = \LK_{S}/(\LK_{S} +
\LK_{q\bar q})$ requirements are determined to maximize
$N_S/\sqrt{N_B}$, with $N_S$ ($N_B$) the expected number of signal (background),
on Monte Carlo (MC) events in each $r$ region separately. More stringent
selections are imposed for the first $r$ region at zero while looser
criteria are used for $r$ close to one. 
More stringent selections are applied for 
decays with large continuum contribution such as \bteprho , while relatively
clean decays such as \btepphi{} have very loose requirements. The signal
efficiencies (continuum background reduction) lie in the range of $42$\%-$88$\%
($98$\%-$45$\%).

Contributions from other charmless $B$
decays can contaminate
the signal when a pion is misidentified as a kaon or when a random
pion is added or missed. 
The dominant contribution for such misidentified events originates from \btepk{}
decays. For the decays \bteprho, \btepkstar{} and \btepomega{} the \btepk{}
contamination is significant. For these decays we construct an alternative $B$
meson hypothesis assuming that it originates from a \btepk{} decay. 
We then veto an event if
the alternative \de{} variable is within a decay-dependent window around $\de =
0$ GeV and $\mb > 5.27$ GeV/$c$. 
The selection is optimized for each decay and results in
negligible signal suppression ($<0.5$\%)
while removing around 80\% of the \btepk{} background. 

\section{Measurement of branching fractions}
The branching fractions are obtained using an extended unbinned
maximum-likelihood fit to the \de{} and \mb{} distributions of selected events.
This fit is performed 
simultaneously in the \epp{} and \erg{} subdecay channels for all 
$B$ decay modes, where applicable. 
In the case of \btepkstarp{} the two $K^{*+}$
subdecay modes (thus four subdecay
channels in total) are fitted simultaneously.
The extended likelihood function used is:
\begin{eqnarray}
L(N_S,N_{B_j}) &=&  \frac{e^{-(N_S+\sum_jN_{B_j})}}{N!} 
\prod_{i=1}^{N}\biggl[N_{S} P_S(\de_i,M_{\mbox{\scriptsize bc}_i})  \nonumber \\
& & + \sum_j N_{B_j} P_{B_j}(\de_i,M_{\mbox{\scriptsize bc}_i})\biggr] ,
\label{eq:ns-lkhd}
\end{eqnarray}
where $N_S$ ($N_{B_j}$) is the number of signal events (background events of
source $j$) with probability density functions (PDFs) 
$P_S$ ($P_{B_j}$). The index 
$i$ runs from 1 to the total number of events $N$ in the selected sample.

The branching fraction $\BF$ is determined by maximizing the 
combined likelihood for both data sets and all subdecays with 
\BF{} constrained to be the same for the subdecays.
The number of signal events ($N_S$) for each decay mode is calculated by
$N_S = \BF \,[ N_{\BB}(I)\, \eff_{\text t}(I)+N_{\BB}(II)\, 
\eff_{\text t}(II) ]$,
where
$N_{\BB}(k)$ is the number of $\BB$ produced for set $k=I$ or $II$ and
$\eff_{\text t}(k)$ is the total reconstruction efficiency including subdecay
branching fractions for set $k$.

The reconstruction efficiencies are determined from signal MC samples
using the EvtGen package~\cite{bib:Evtgen} with final
state radiation simulated by the PHOTOS package~\cite{bib:Photos} (thus
measuring \btephh $(\gamma)$).
The efficiencies are calculated separately for
Set $I$ and Set $II$. The absolute
efficiency for Set $II$ is typically about 0.5\%
larger than for Set $I$ (for efficiencies averaged over the two sets 
see Tables~\ref{tab:results1} and~\ref{tab:results2}).
Corrections due to differences
between data and MC are included for the charged track
identification and photon, $\pi^0$ and $\eta$ reconstructions, 
resulting in an overall correction factor between $0.88$ and $0.99$
depending on the decay mode. We assume the numbers of $B^+B^-$ and $B^0\Bbaro$
pairs to be equal in the original data sample.

In the fit to the data we consider a signal component and three types of
background components: continuum events, events from other $B$ meson decays via
the dominant $b\to c$ transition and from charmless $B$ decays. 

For both signal and continuum background \de{} and \mb{} are uncorrelated and we
use two independent functions to describe the shapes of \de{} and \mb.
To model the signal, we use a Gaussian with an
exponential tail, the so-called Crystal Ball-line (CBline)
function~\cite{Gaiser:1986ix}, plus a Gaussian in \de , while \mb{} 
is described by
a single CBline function. 
The shape parameters are fixed from the signal MC study. 
Corrections for MC-data discrepancies determined
from control samples of \btepkp{} and $B^+\to D^0 \pi^0$, where $D^0 \to K^-
\pi^+$  and $D^0 \to K^- \pi^+ \pi^0$, 
are applied to the mean and width of
the CBline functions.

Continuum background is modelled by a first-order 
polynomial for \de{} and an ARGUS function~\cite{bib:ARGUS} for \mb. 
The continuum shape parameters, that are allowed to float in all modes, are the
slopes of the polynomial and ARGUS function.
The shapes for charmless
$B$ decays remaining after applying the vetoes 
and $b\to c$ backgrounds are modelled by two-dimensional smoothed histograms. 
The sizes of background contributions other than the continuum background
are fixed to 
the values expected from MC studies. 

\begin{figure*}[htb]
\unitlength1.0cm
\hspace{0.29cm}
\framebox[14.18cm][c]{
\textcolor{black}{+ \, data} \hspace{0.5cm}
\textcolor{green}{----- combined} \hspace{0.5cm}
\textcolor{red}{ -\hspace{2pt}-\hspace{2pt}-\hspace{2pt}- signal} \hspace{0.5cm}
\textcolor{violet}{$\cdot \!\! \cdot \!\! \cdot \!\! \cdot \!\!\cdot \!\! \cdot
\!\!\cdot \! \cdot$ udsc} \hspace{0.5cm}
\textcolor{blue}{--  --  -- \BB{}} \hspace{0.5cm}
\textcolor{magenta}{-$\cdot$-$\cdot$ rare B}
}
\includegraphics[width=0.22\textwidth]{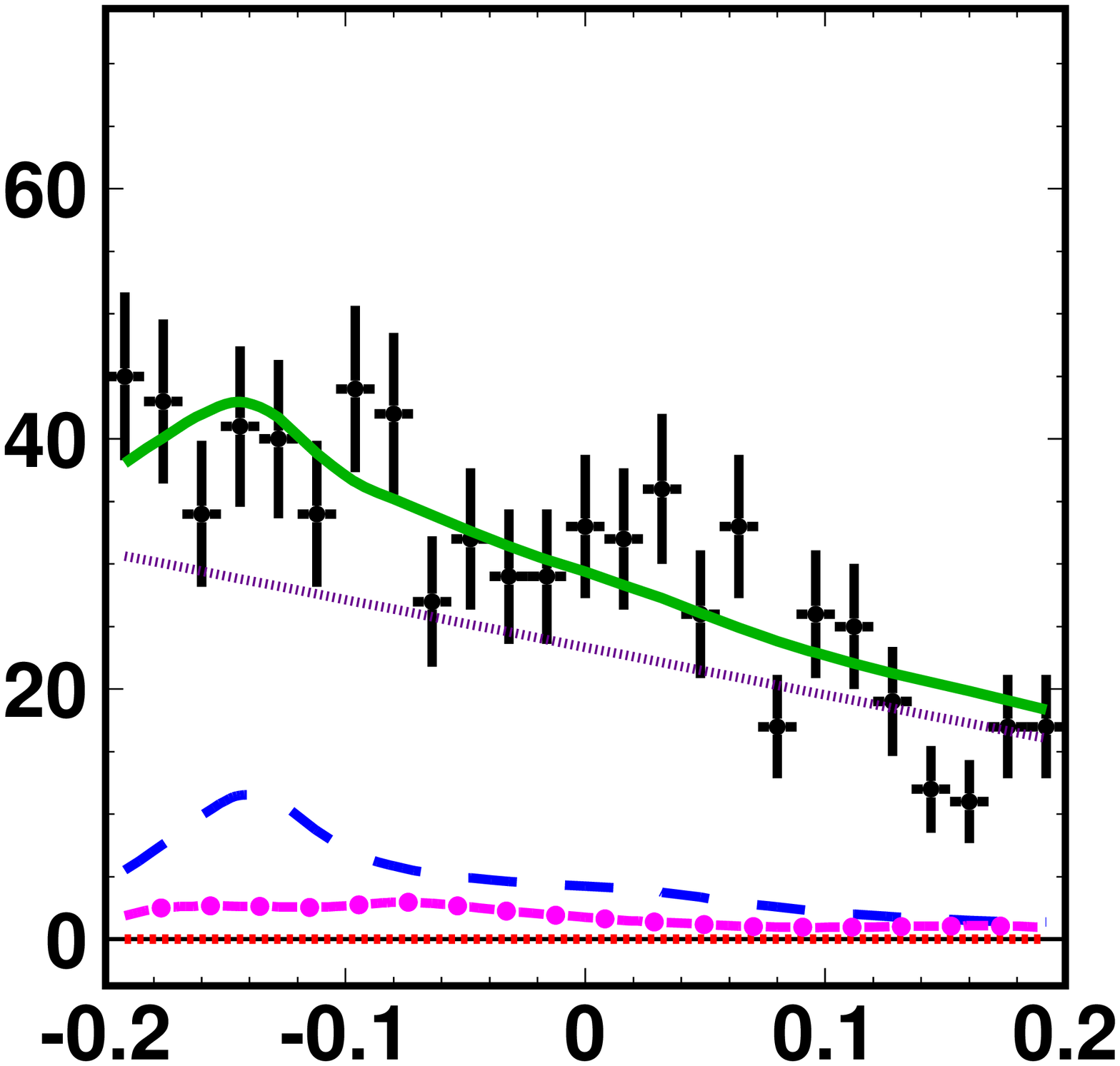}
\includegraphics[width=0.22\textwidth]{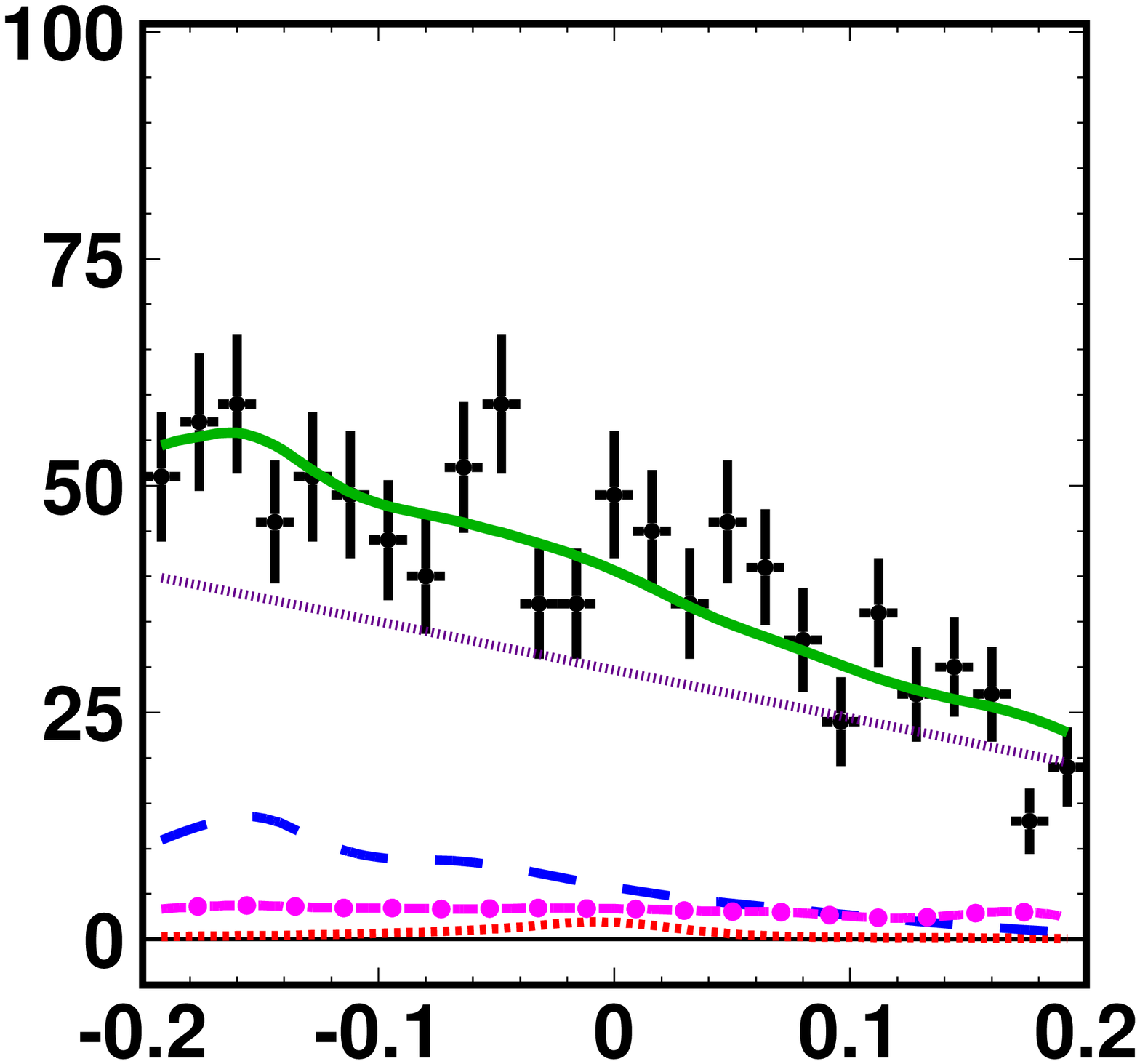}
\includegraphics[width=0.22\textwidth]{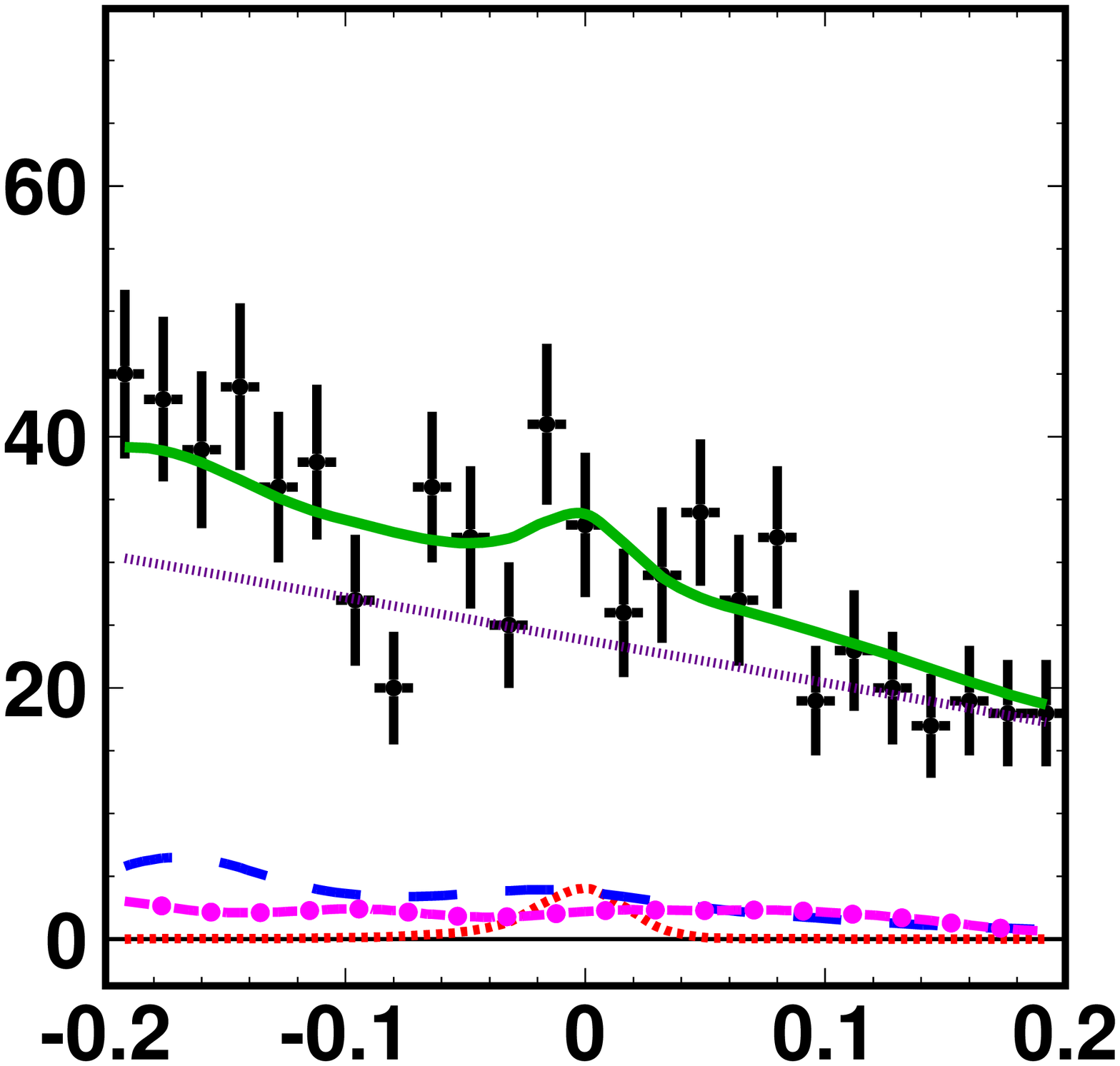}
\includegraphics[width=0.22\textwidth]{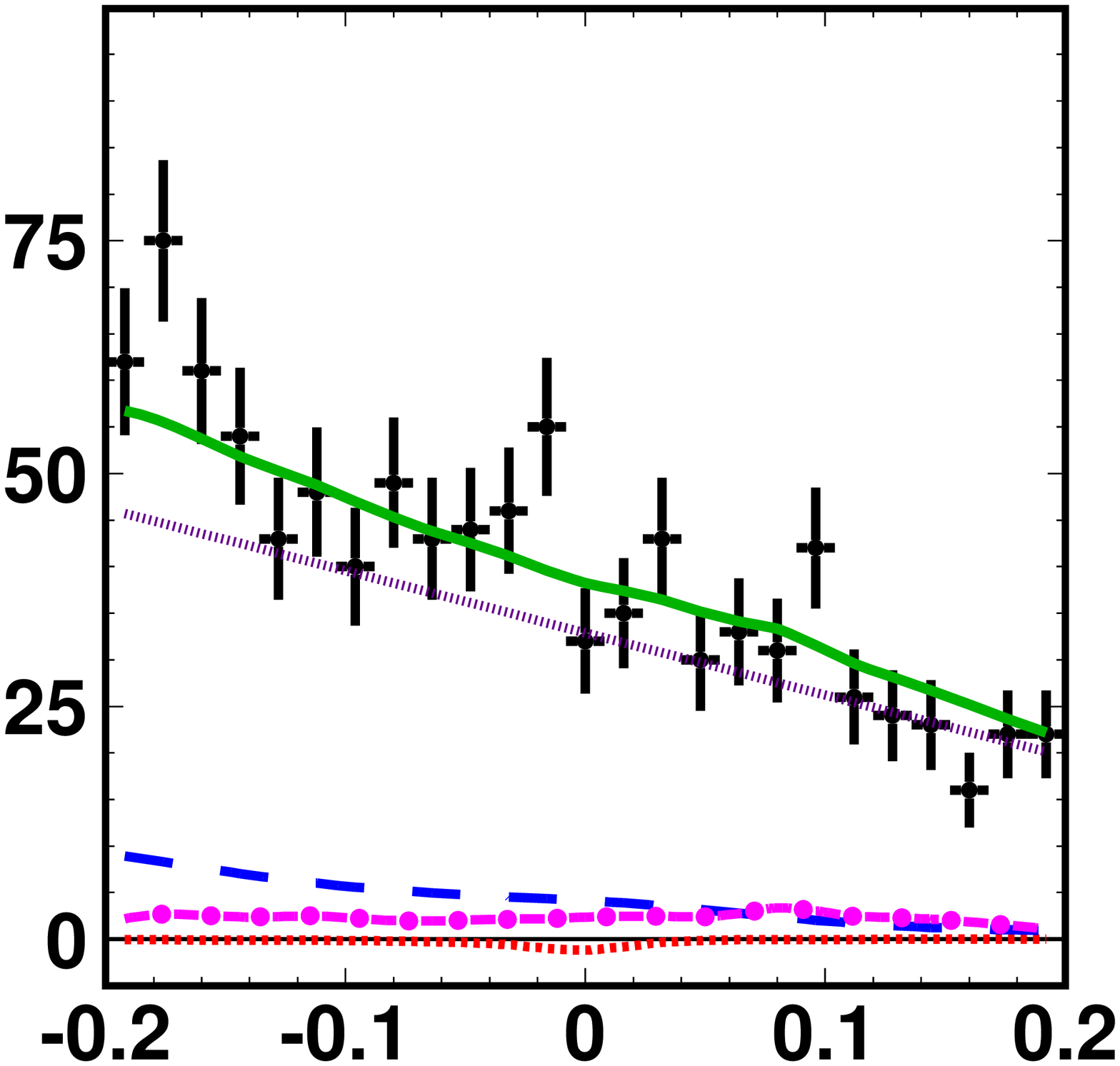}
\includegraphics[width=0.22\textwidth]{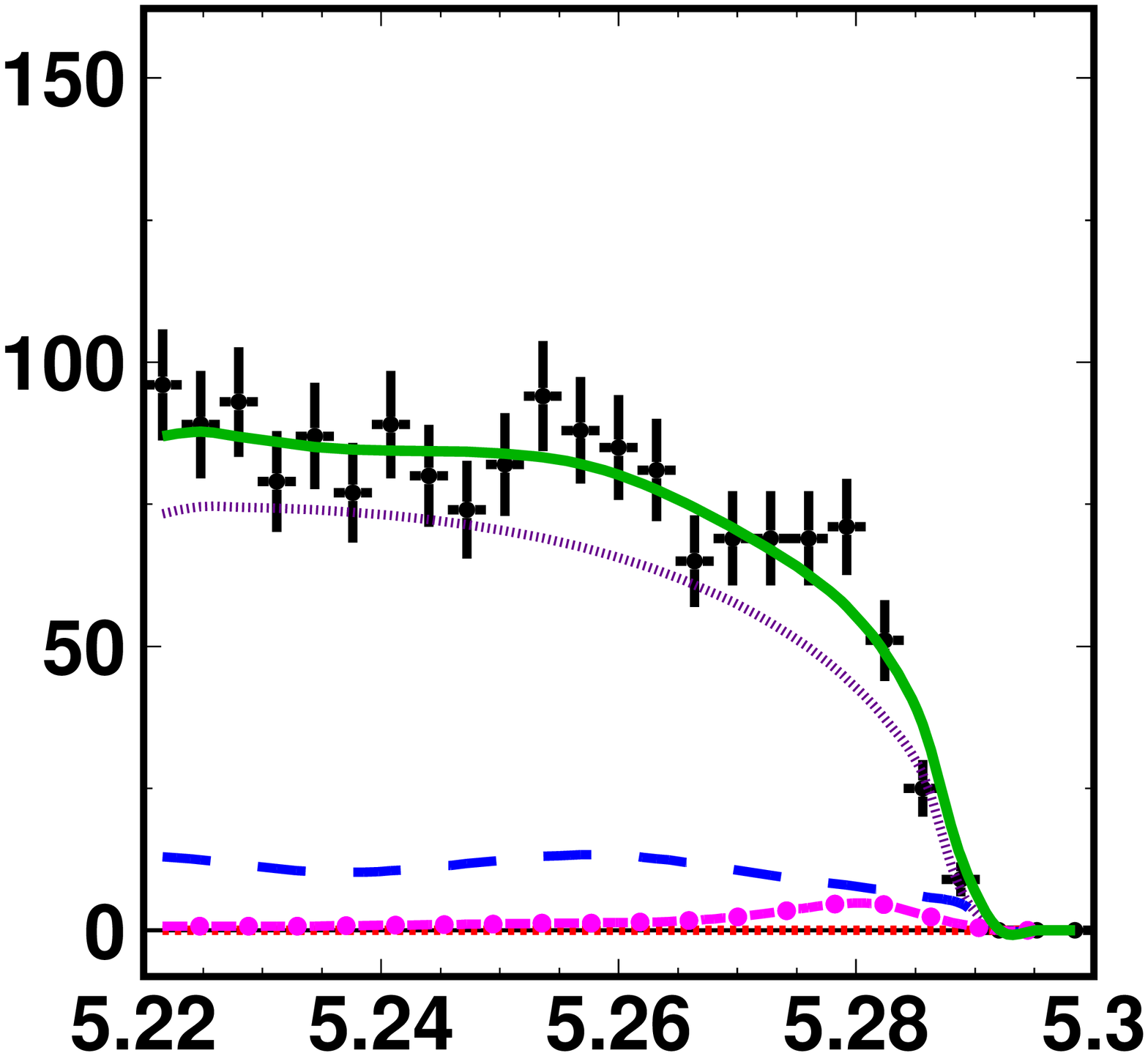}
\includegraphics[width=0.22\textwidth]{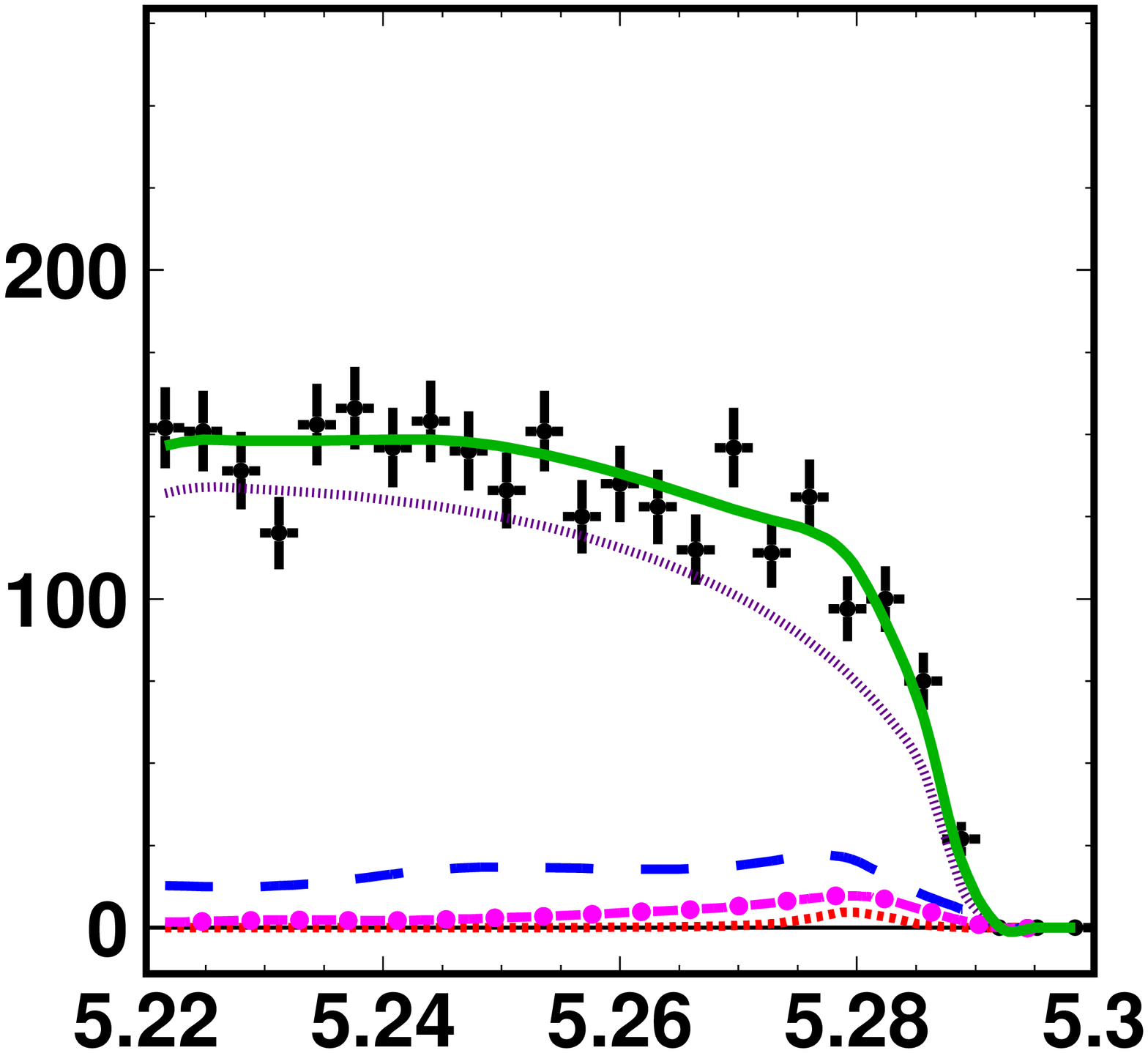}
\includegraphics[width=0.22\textwidth]{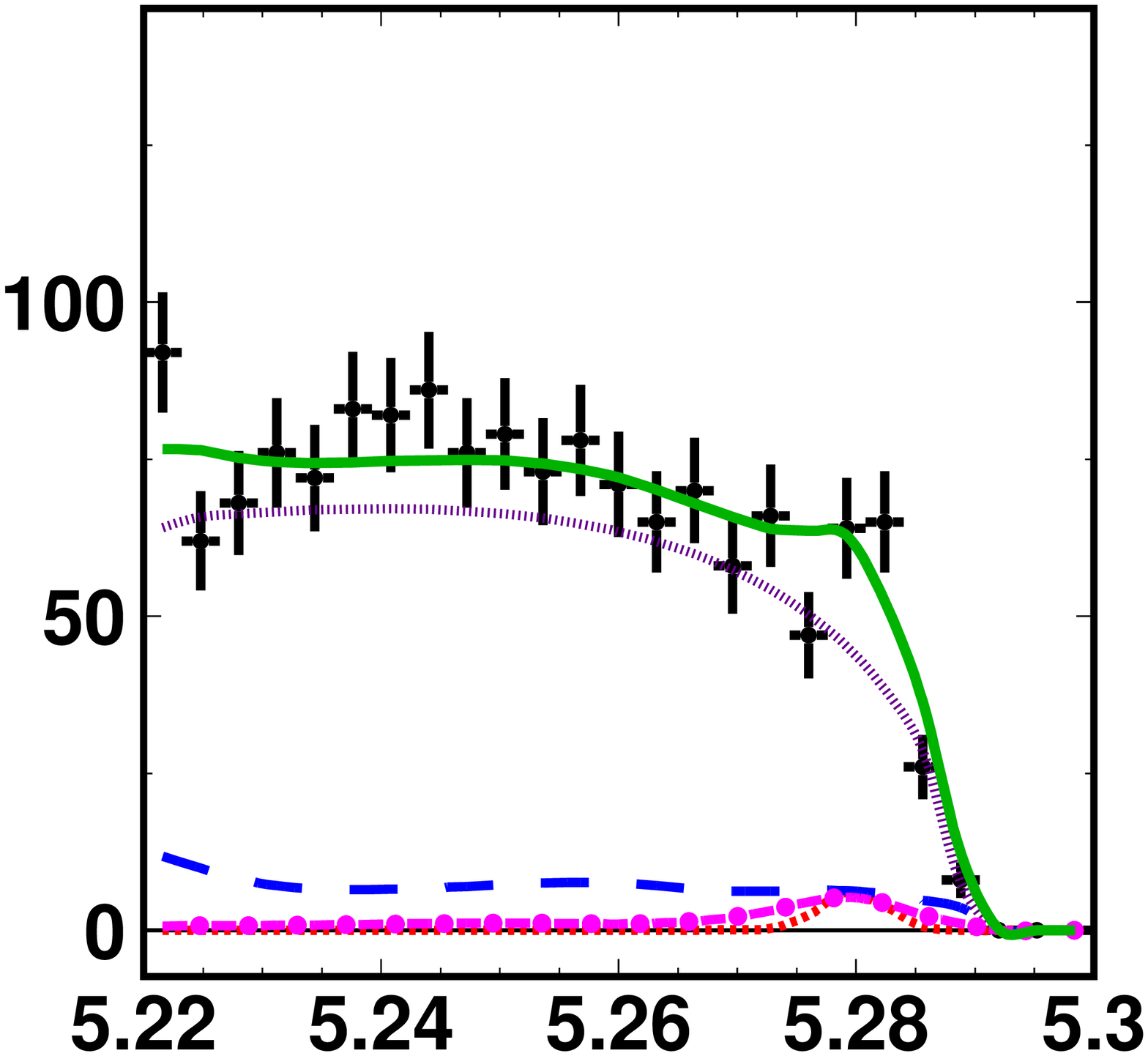}
\includegraphics[width=0.22\textwidth]{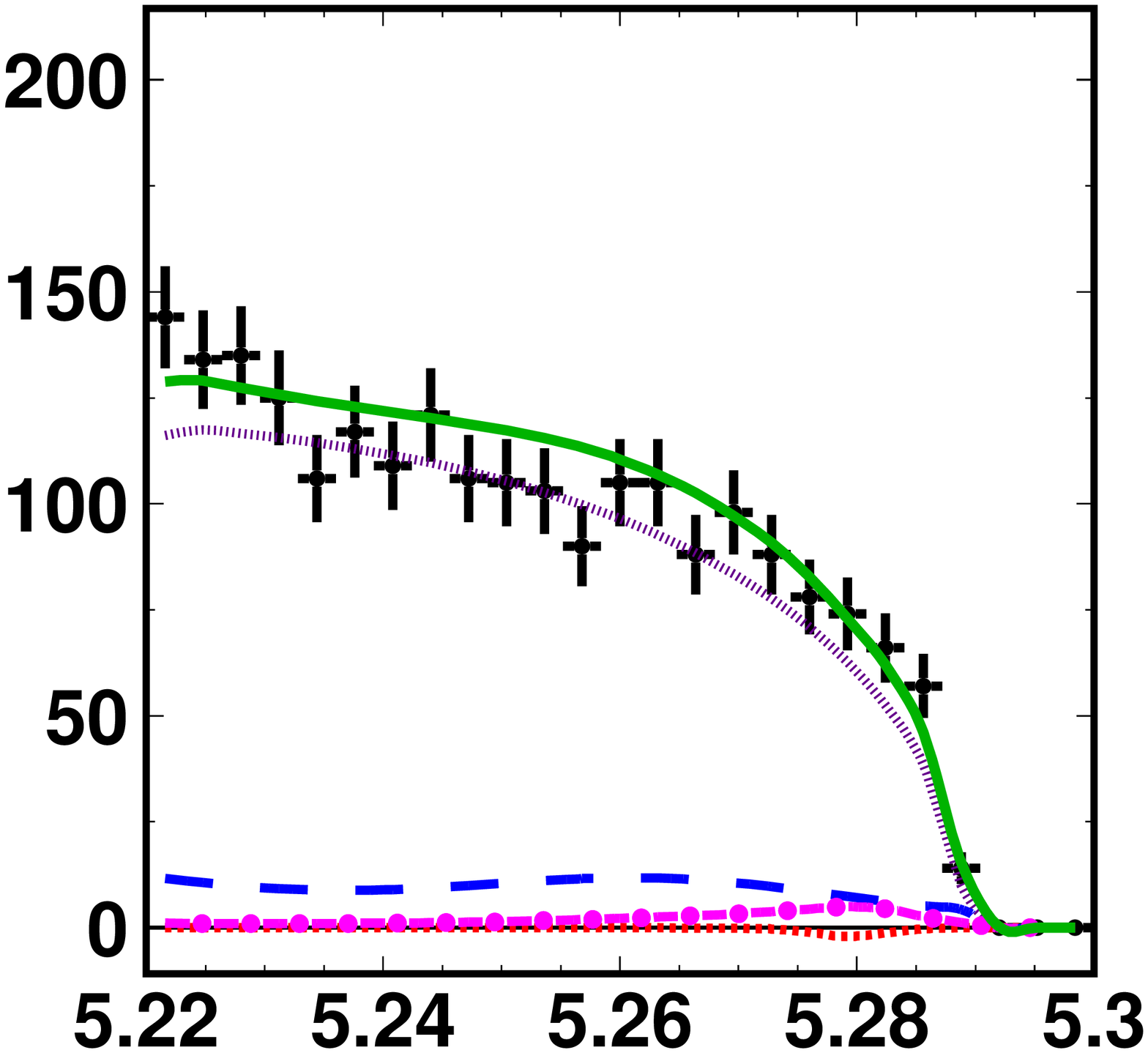}
\begin{rotate}{90}
\put(0.5,15.2){{\footnotesize{\sf\shortstack[c]{{Entries / 3.2 Mev\cs}}}}}
\put(5.3,15.2){{\footnotesize{\sf\shortstack[c]{{Entries / 16 Mev\cs}}}}}
\end{rotate}
\put(-14.1,3.9){\large{\sf\shortstack[c]{\bteprhoo}}}
\put(-10.3,3.9){\large{\sf\shortstack[c]{\bteprhop}}}
\put(-6.65,3.9){\large{\sf\shortstack[c]{\btepkstaro}}}
\put(-2.8,3.9){\large{\sf\shortstack[c]{\btepkstarp}}}
\put(-13.8,4.5){\footnotesize{\sf\shortstack[c]{\de{} [GeV]}}}
\put(-10.0,4.5){\footnotesize{\sf\shortstack[c]{\de{} [GeV]}}}
\put(-6.35,4.5){\footnotesize{\sf\shortstack[c]{\de{} [GeV]}}}
\put(-2.5,4.5){\footnotesize{\sf\shortstack[c]{\de{} [GeV]}}}
\put(-13.8,-0.2){\footnotesize{\sf\shortstack[c]{\mb{} [GeV/$c$]}}}
\put(-10.0,-0.2){\footnotesize{\sf\shortstack[c]{\mb{} [GeV/$c$]}}}
\put(-6.35,-0.2){\footnotesize{\sf\shortstack[c]{\mb{} [GeV/$c$]}}}
\put(-2.5,-0.2){\footnotesize{\sf\shortstack[c]{\mb{} [GeV/$c$]}}}
\caption{\de{} (upper) and \mb{} (lower) distributions for (from left to right)
\bteprhoo ,
\bteprhop, \btepkstaro{} and \btepkstarp.}
\label{fig:rho}
\end{figure*}

\begin{figure*}[htb]
\unitlength1.0cm
\unitlength1.0cm
\hspace{0.28cm}
\framebox[14.18cm][c]{
\textcolor{black}{+ \, data} \hspace{0.5cm}
\textcolor{green}{----- combined} \hspace{0.5cm}
\textcolor{red}{ -\hspace{2pt}-\hspace{2pt}-\hspace{2pt}- signal} \hspace{0.5cm}
\textcolor{violet}{$\cdot \!\! \cdot \!\! \cdot \!\! \cdot \!\!\cdot \!\! \cdot
\!\!\cdot \! \cdot$ udsc} \hspace{0.5cm}
\textcolor{blue}{--  --  -- \BB{}} \hspace{0.5cm}
\textcolor{magenta}{-$\cdot$-$\cdot$ rare B}
}
\includegraphics[width=0.22\textwidth]{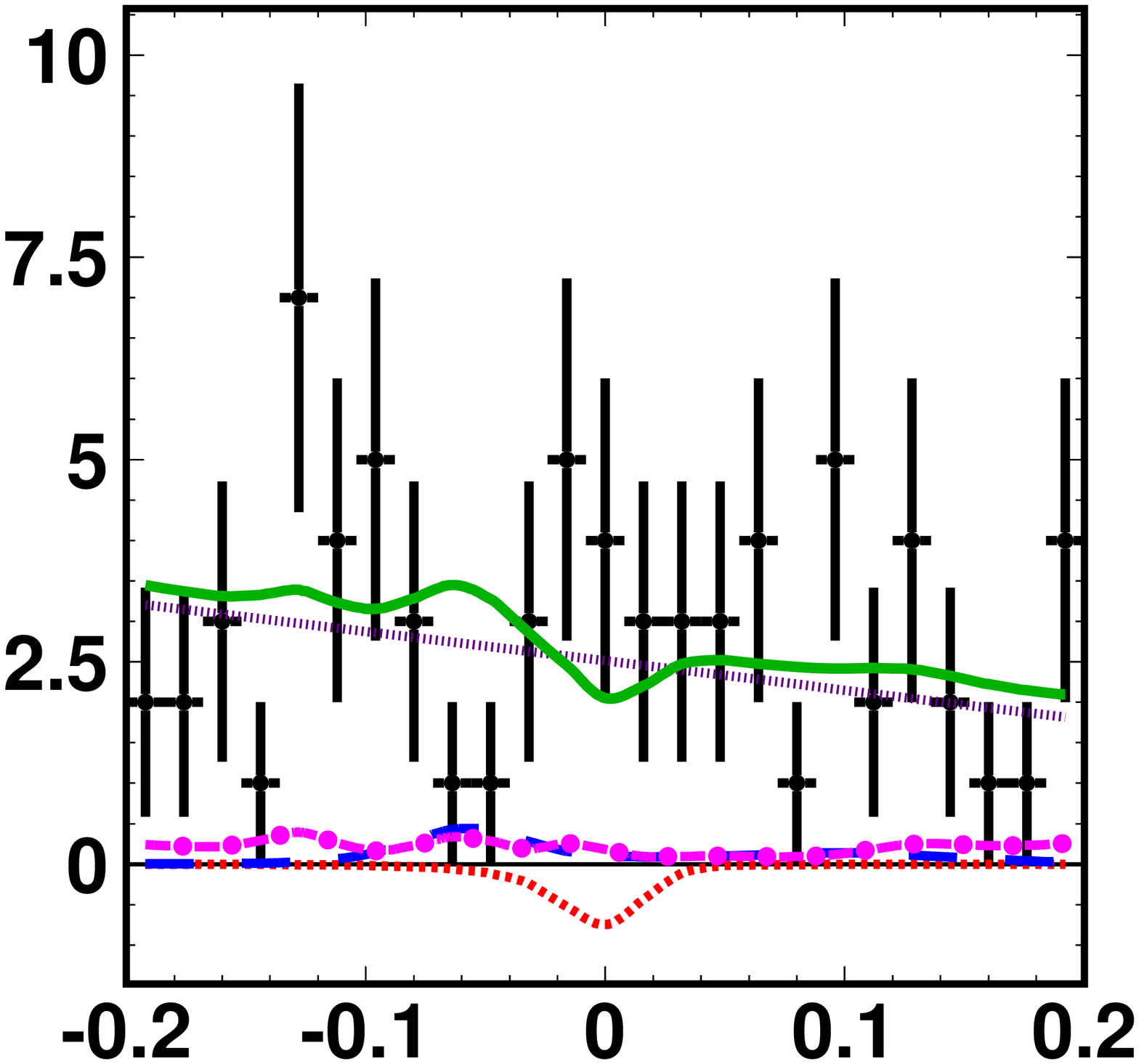}
\includegraphics[width=0.22\textwidth]{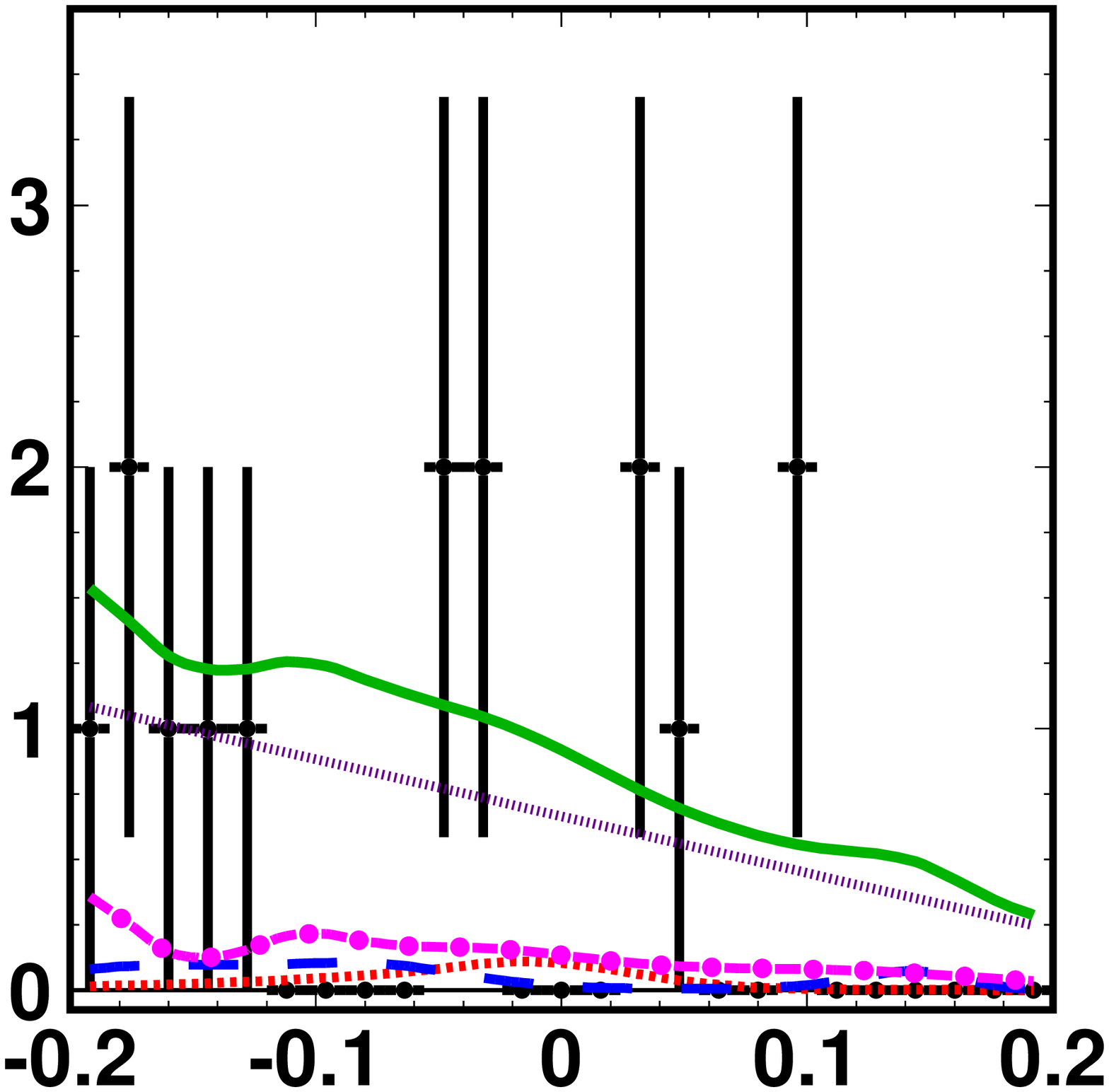}
\includegraphics[width=0.22\textwidth]{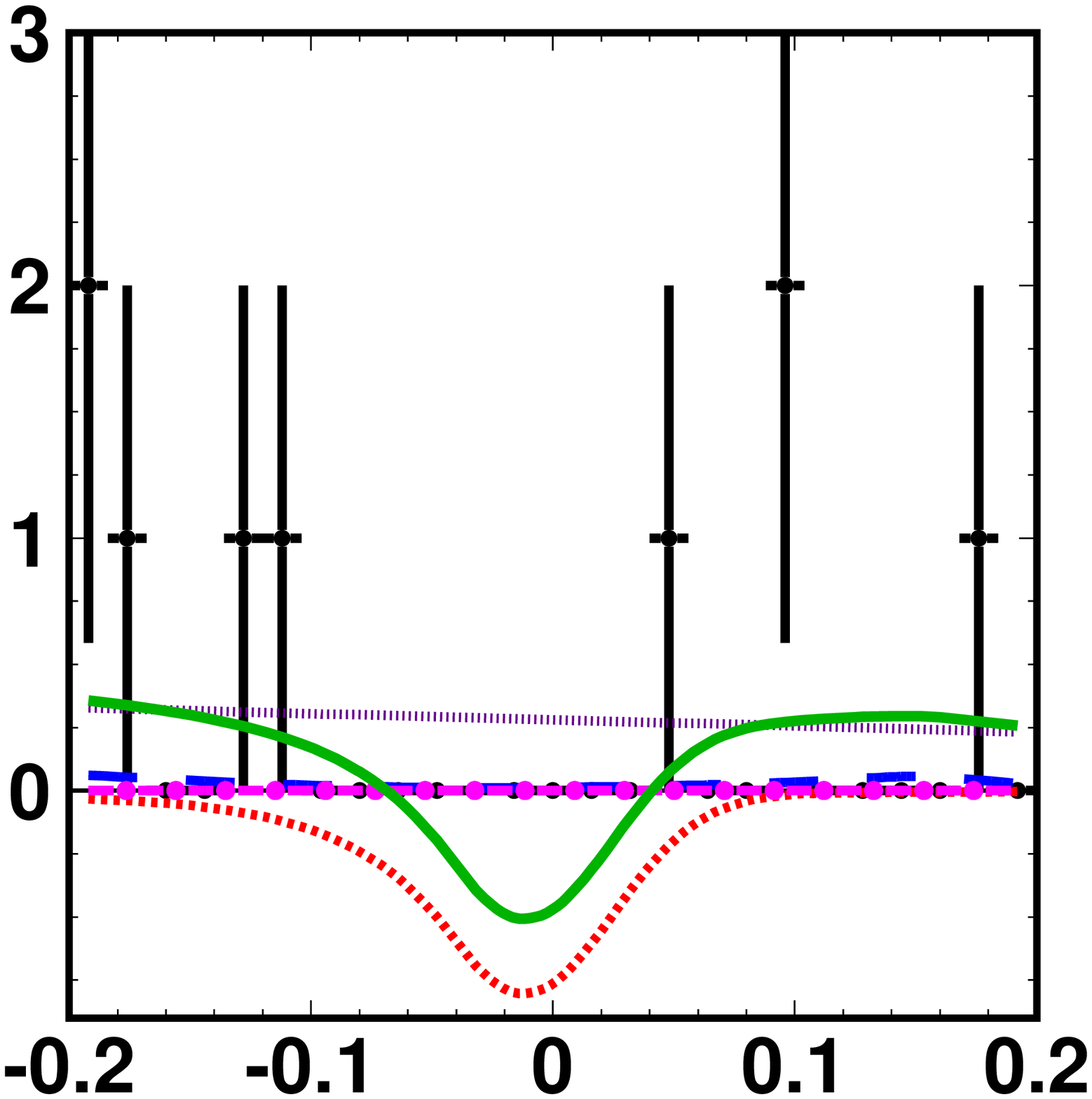}
\includegraphics[width=0.22\textwidth]{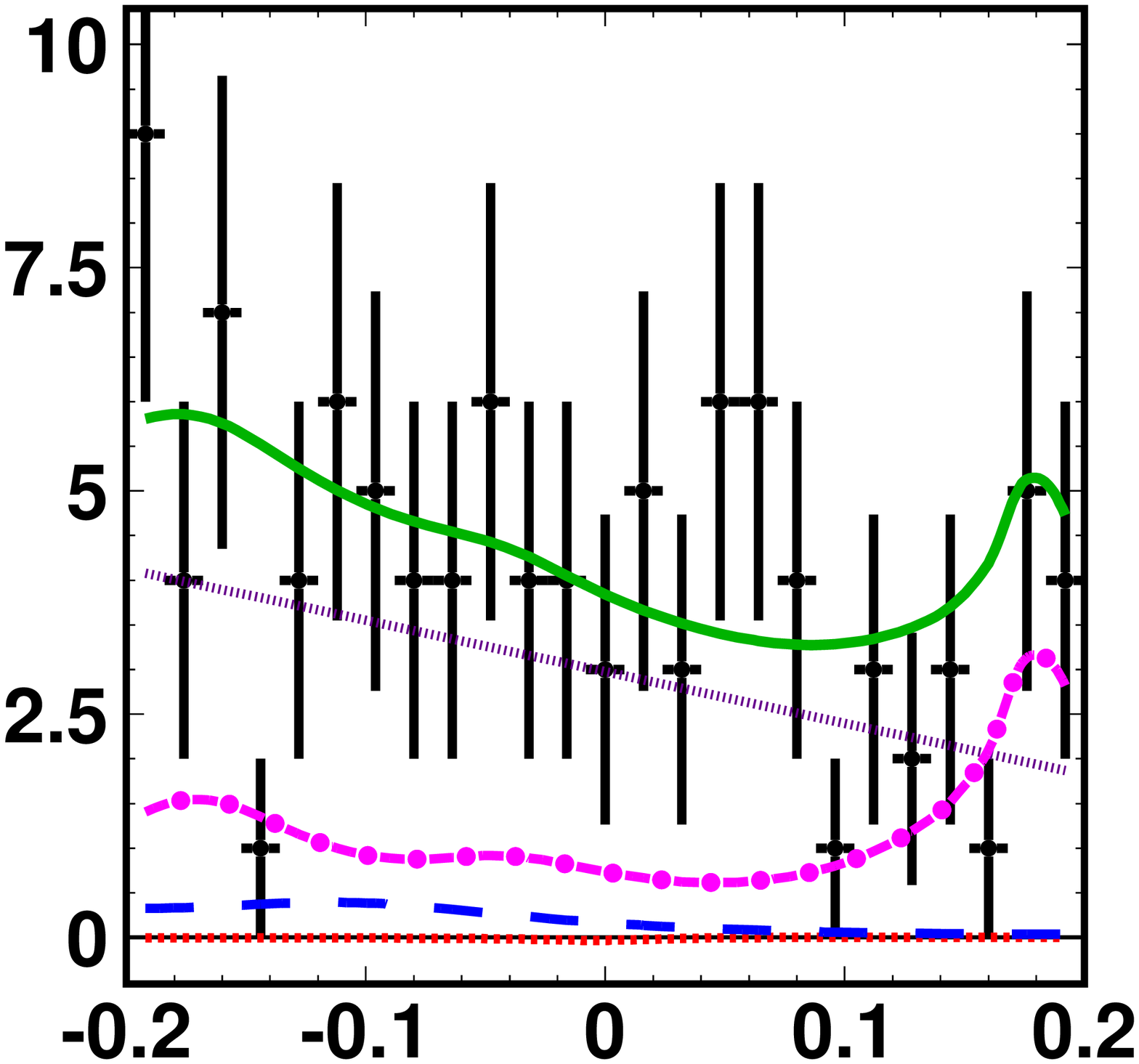}
\includegraphics[width=0.22\textwidth]{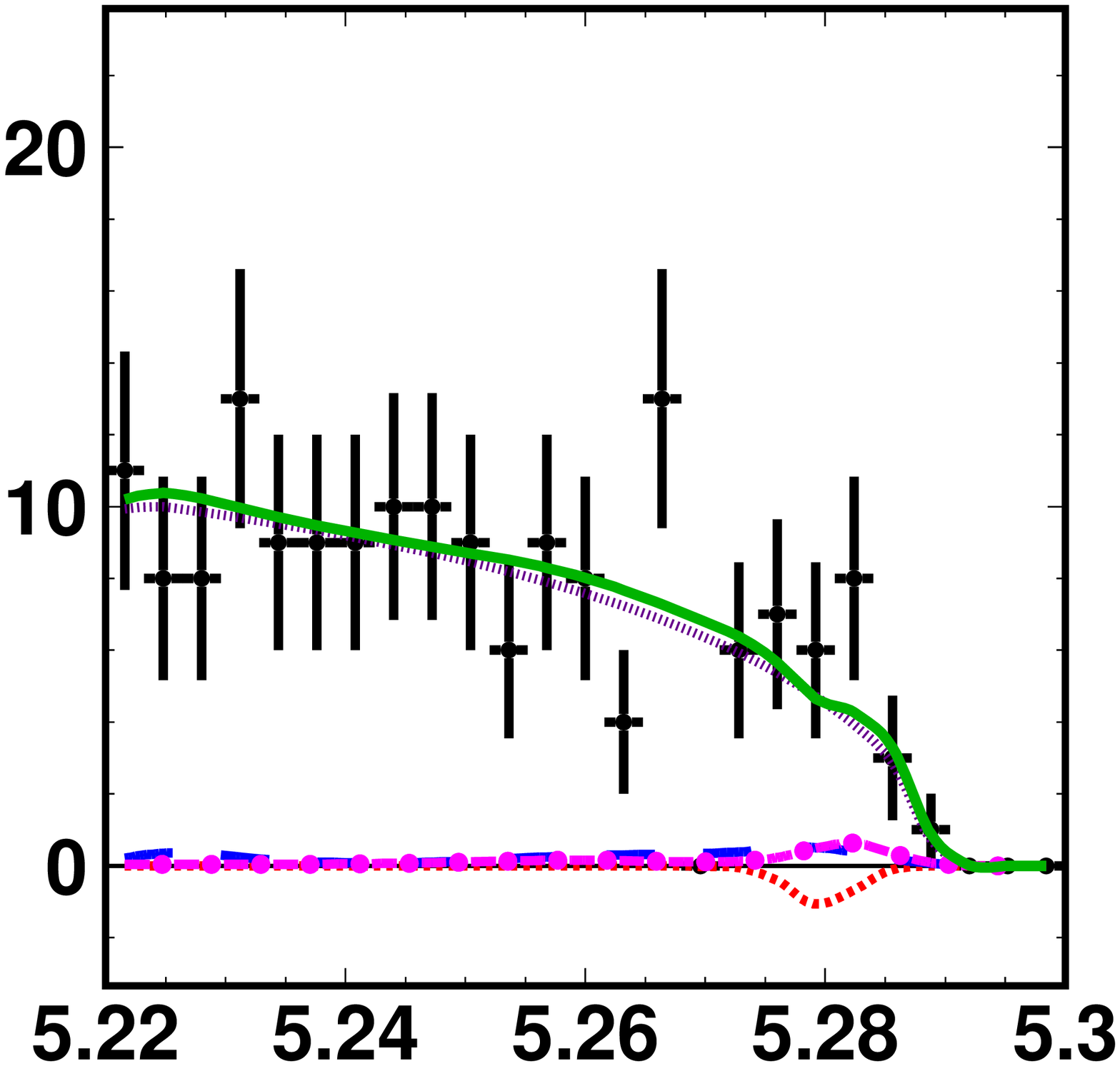}
\includegraphics[width=0.22\textwidth]{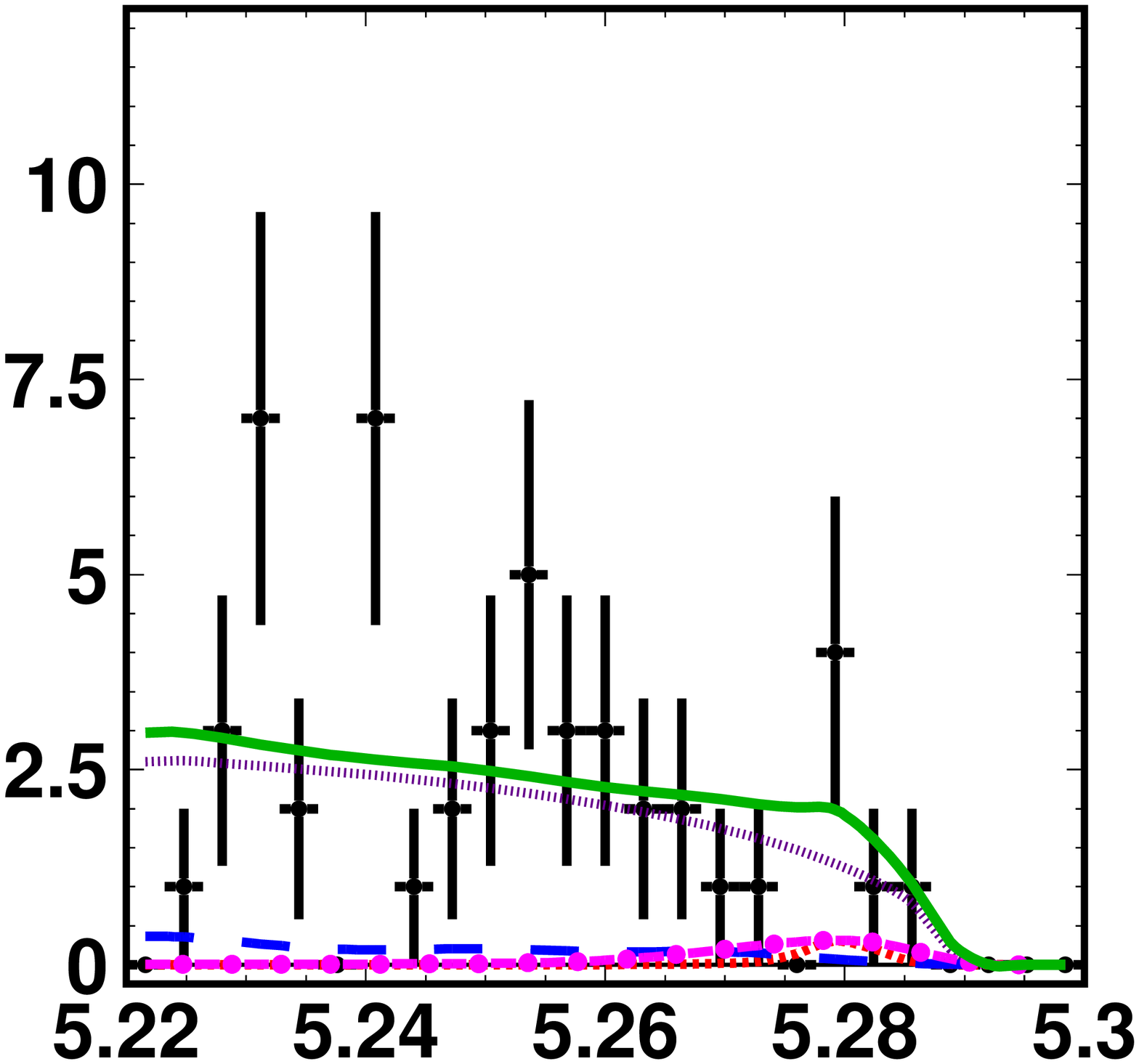}
\includegraphics[width=0.22\textwidth]{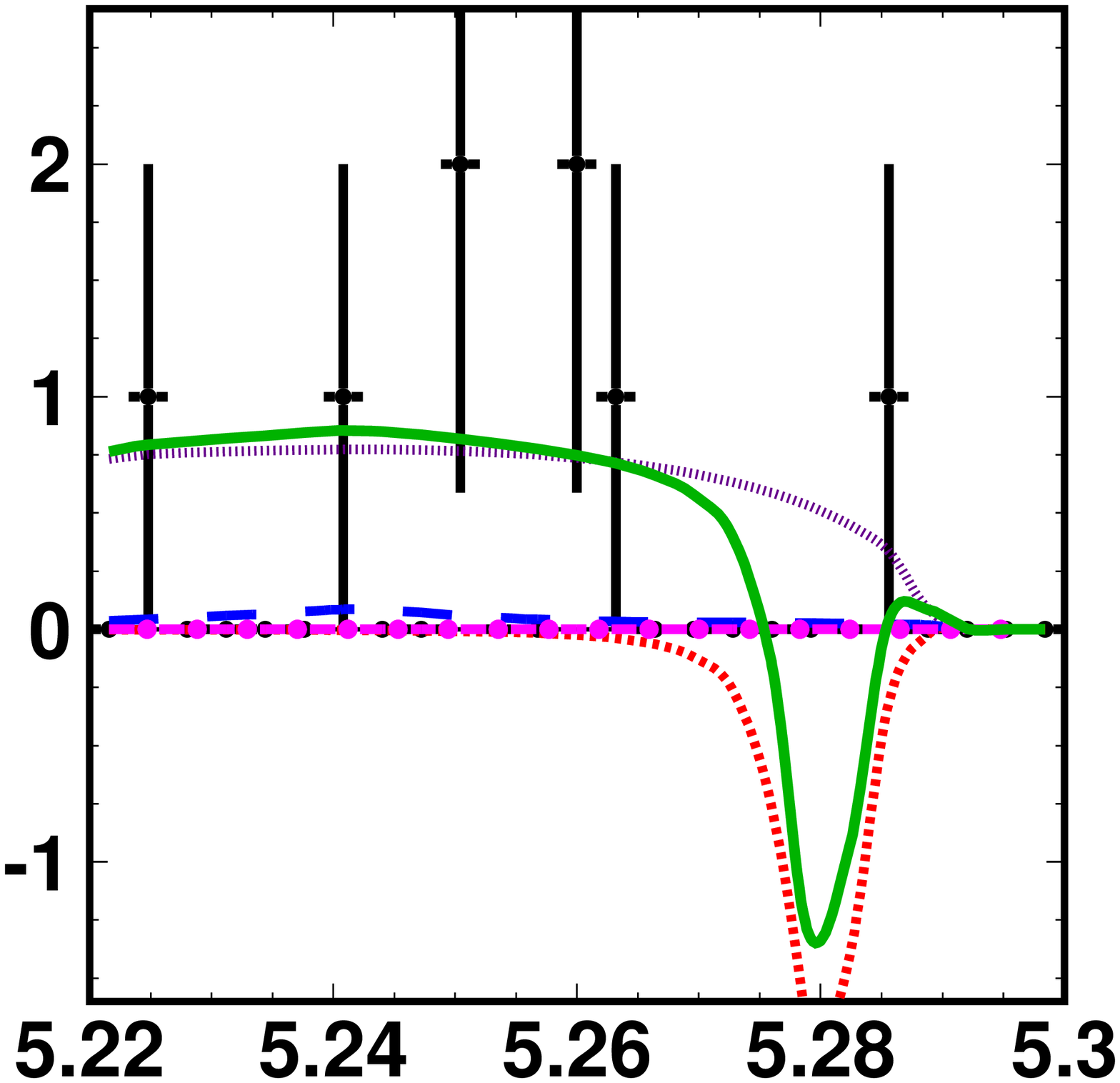}
\includegraphics[width=0.22\textwidth]{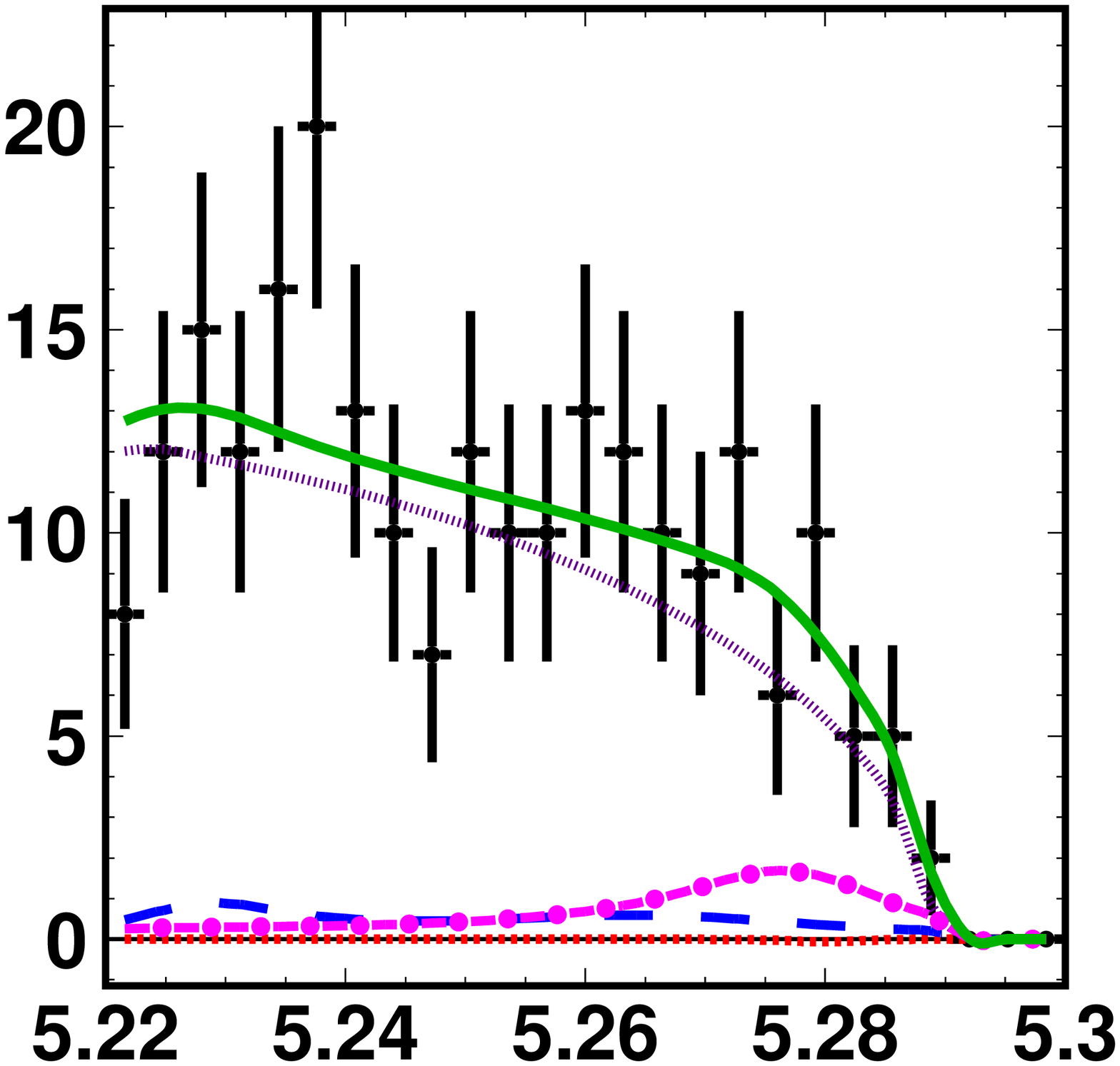}
\begin{rotate}{90}
\put(0.5,15.2){{\footnotesize{\sf\shortstack[c]{{Entries / 3.2 Mev\cs}}}}}
\put(5.3,15.2){{\footnotesize{\sf\shortstack[c]{{Entries / 16 Mev\cs}}}}}
\end{rotate}
\put(-14.1,3.9){\large{\sf\shortstack[c]{\btepphi}}}
\put(-10.3,3.9){\large{\sf\shortstack[c]{\btepeta}}}
\put(-6.65,3.9){\large{\sf\shortstack[c]{\btepetap}}}
\put(-2.8,3.9){\large{\sf\shortstack[c]{\btepomega}}}
\put(-13.8,4.5){\footnotesize{\sf\shortstack[c]{\de{} [GeV]}}}
\put(-10.0,4.5){\footnotesize{\sf\shortstack[c]{\de{} [GeV]}}}
\put(-6.35,4.5){\footnotesize{\sf\shortstack[c]{\de{} [GeV]}}}
\put(-2.5,4.5){\footnotesize{\sf\shortstack[c]{\de{} [GeV]}}}
\put(-13.8,-0.2){\footnotesize{\sf\shortstack[c]{\mb{} [GeV/$c$]}}}
\put(-10.0,-0.2){\footnotesize{\sf\shortstack[c]{\mb{} [GeV/$c$]}}}
\put(-6.35,-0.2){\footnotesize{\sf\shortstack[c]{\mb{} [GeV/$c$]}}}
\put(-2.5,-0.2){\footnotesize{\sf\shortstack[c]{\mb{} [GeV/$c$]}}}
\caption{\de{} (upper) and \mb{} (lower) distributions for (from left to right)
\btepphi , \btepeta ,
\btepetap{} and \btepomega.}
\label{fig:phi}
\end{figure*}

The resulting \de{} and \mb{} projections are shown in
Figs.~\ref{fig:rho} and~\ref{fig:phi}.
The reconstruction efficiencies and fit results are given in 
Tables~\ref{tab:results1} and~\ref{tab:results2}. 
\begin{table*}[htb]
\caption{Average efficiencies ($\epsilon$) 
for the two data sets
for \epp{} and \erg , total efficiencies ($\epsilon_{\text t}$) with 
systematic errors of secondary branching fractions included,
signal yield ($N_S$) with statistical errors only
and the 90\% confidence level upper limit on the branching fraction in units of
$10^{-6}$ including systematic errors
for each decay of this analysis (UL) and latest results from \babar{}
in units of $10^{-6}$.}
\label{tab:results1}
\begin{tabular}
{@{\hspace{0.4cm}}l@{\hspace{0.0cm}}@{\hspace{0.0cm}}c@{\hspace{0.3cm}}
@{\hspace{0.3cm}}c@{\hspace{0.3cm}}@{\hspace{0.3cm}}c@{\hspace{0.3cm}}
@{\hspace{0.3cm}}c@{\hspace{0.3cm}}@{\hspace{0.3cm}}c@{\hspace{0.3cm}}
@{\hspace{0.3cm}}c@{\hspace{0.3cm}}}
\hline \hline
  & & \bteprhoo 	& \bteprhop 	& \btepkstaro 	& \btepkstarp	& \btepphi \\
\hline
$\epsilon (\nepp )$ &[\%] 		& $7.0\pm 0.1$ 		& $5.9\pm 0.1$ 		& $8.5\pm 0.1$		& $4.5\pm 0.1$			& $12.9\pm 0.1$		\\
$\epsilon (\nerg )$ &[\%] 		& $5.4\pm 0.1$ 		& $3.9\pm 0.1$ 		& $5.9\pm 0.1$ 		& $2.2\pm 0.1$ 			& $7.4\pm 0.1$		\\
$\epsilon_{\text t}(\nepp )$ &[\%] 	& $1.13\pm 0.02$ 	& $0.93\pm 0.02$  	& $0.92\pm 0.01$	& $0.35\pm 0.01$		& $1.08\pm 0.01$	\\
$\epsilon_{\text t}(\nerg )$ &[\%] 	& $1.51\pm 0.03$ 	& $1.07\pm 0.02$	& $1.09\pm 0.02$	& $0.30\pm 0.01$		& $1.08 \pm 0.02$	\\
$N_S$ 		&			& $0.1^{+8.2}_{-7.0}$ 	& $18.5^{+23.3}_{-21.7}$ & $14.2^{+9.1}_{-8.0}$	& $-6.4^{+10.9}_{-7.9}$		& $-2.4^{+2.5}_{-3.5}$	\\
UL & [$10^{-6}$]				& $ <1.3$ 		& $<5.8$		& $<2.6$		& $<2.9$			& $<0.5$		\\
\babar{} & [$10^{-6}$]			& $< 3.7$ 		& $< 14 $		& $3.8 \pm 1.1 \pm \! 0.5$ & $4.9^{+1.9}_{-1.7}\pm \! 0.8$	&  $< 4.5$ 	\\
\hline \hline
\end{tabular}
\end{table*}
\begin{table*}[htb]
\caption{Average efficiencies ($\epsilon$) 
for the two data sets
for \epp , total efficiencies ($\epsilon_{\text t}$) with systematic
errors of secondary branching fractions included,
signal yield ($N_S$) with statistical errors only
and the 90\% confidence level upper limit on the branching fraction in units of
$10^{-6}$ including systematic errors
for each decay of this analysis (UL) and latest results from \babar{}
in units of $10^{-6}$.}
\label{tab:results2}
\begin{tabular}
{@{\hspace{0.4cm}}l@{\hspace{0.0cm}}@{\hspace{0.0cm}}c@{\hspace{0.4cm}}
@{\hspace{0.4cm}}c@{\hspace{0.4cm}}@{\hspace{0.4cm}}c@{\hspace{0.4cm}}
@{\hspace{0.4cm}}c@{\hspace{0.4cm}}}
\hline \hline
  		&			& \btepeta		& \btepetap 		& \btepomega  	\\
\hline
$\epsilon (\nepp )$ &[\%]		& $5.7\pm 0.1$		& $4.8\pm 0.1$		& $7.5\pm 0.1$		\\
$\epsilon_{\text t}(\nepp )$ &[\%] 	& $0.37\pm 0.007$	& $0.16\pm 0.003$	& $1.09\pm 0.02$	\\
$N_S$ 			&		& $1.0^{+4.6}_{-3.6}$	& $-6.3^{+2.2}_{-2.1}$	& $0.9^{+6.3}_{-5.2}$	\\
UL  & [$10^{-6}$]				& $< 4.5$		& $< 6.5$		& $< 2.2$		\\
\babar{} & [$10^{-6}$]			& $< 1.7$		& $< 10 $		& $< 2.8$		\\
\hline \hline
\end{tabular}
\end{table*}
\begin{figure*}[htb]
\unitlength1.0cm
\includegraphics[width=0.22\textwidth]{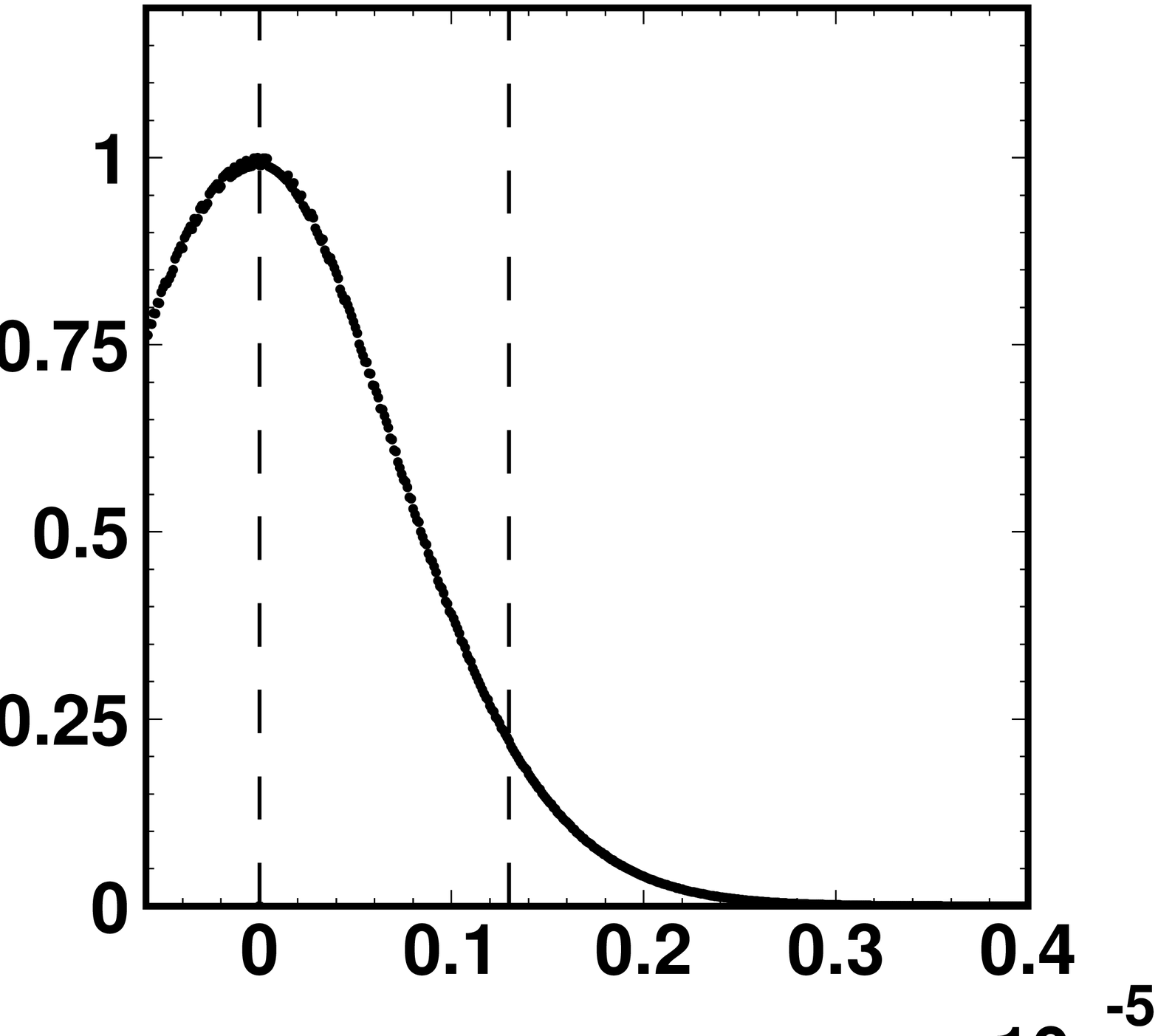}
\includegraphics[width=0.22\textwidth]{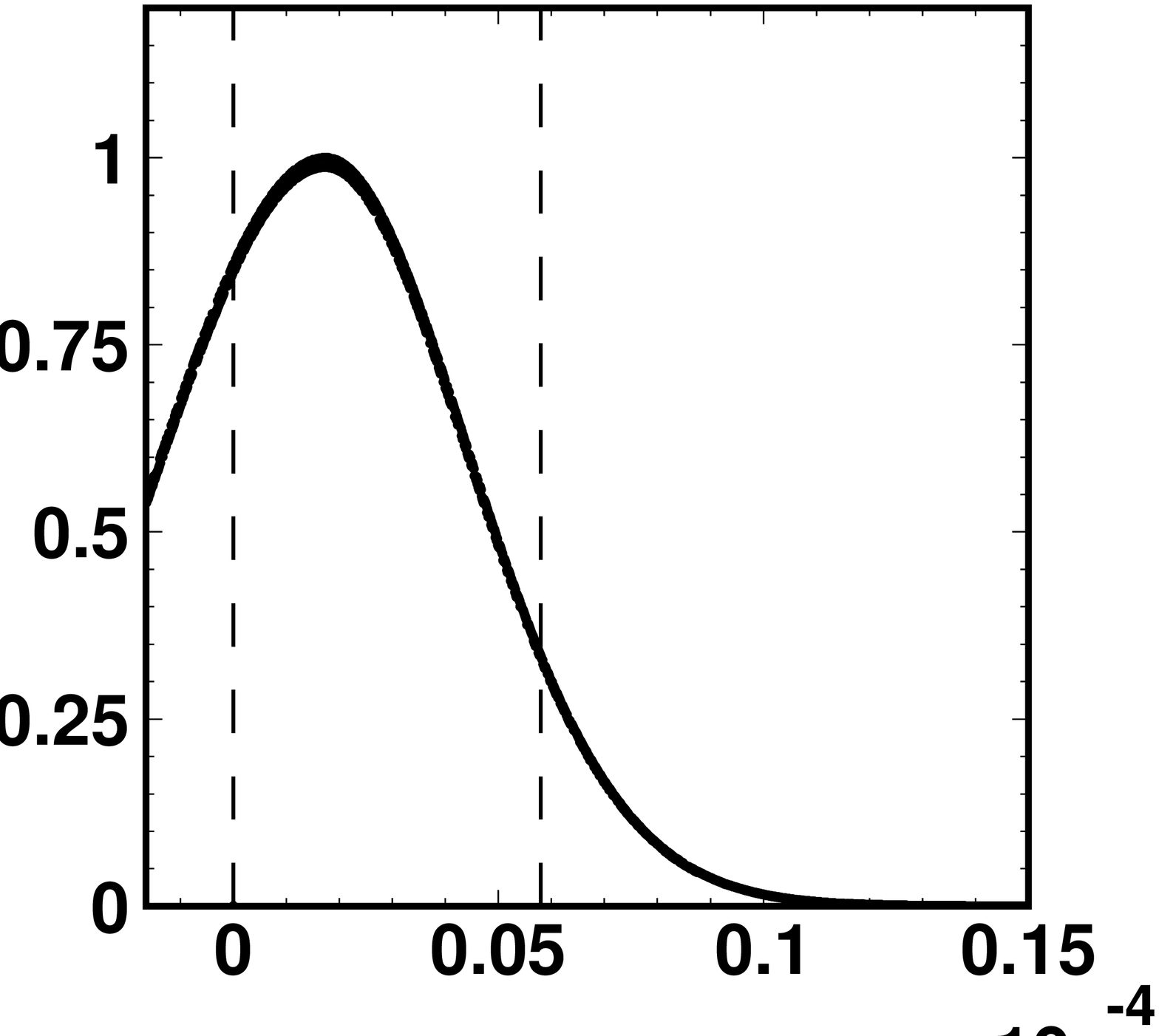}
\includegraphics[width=0.22\textwidth]{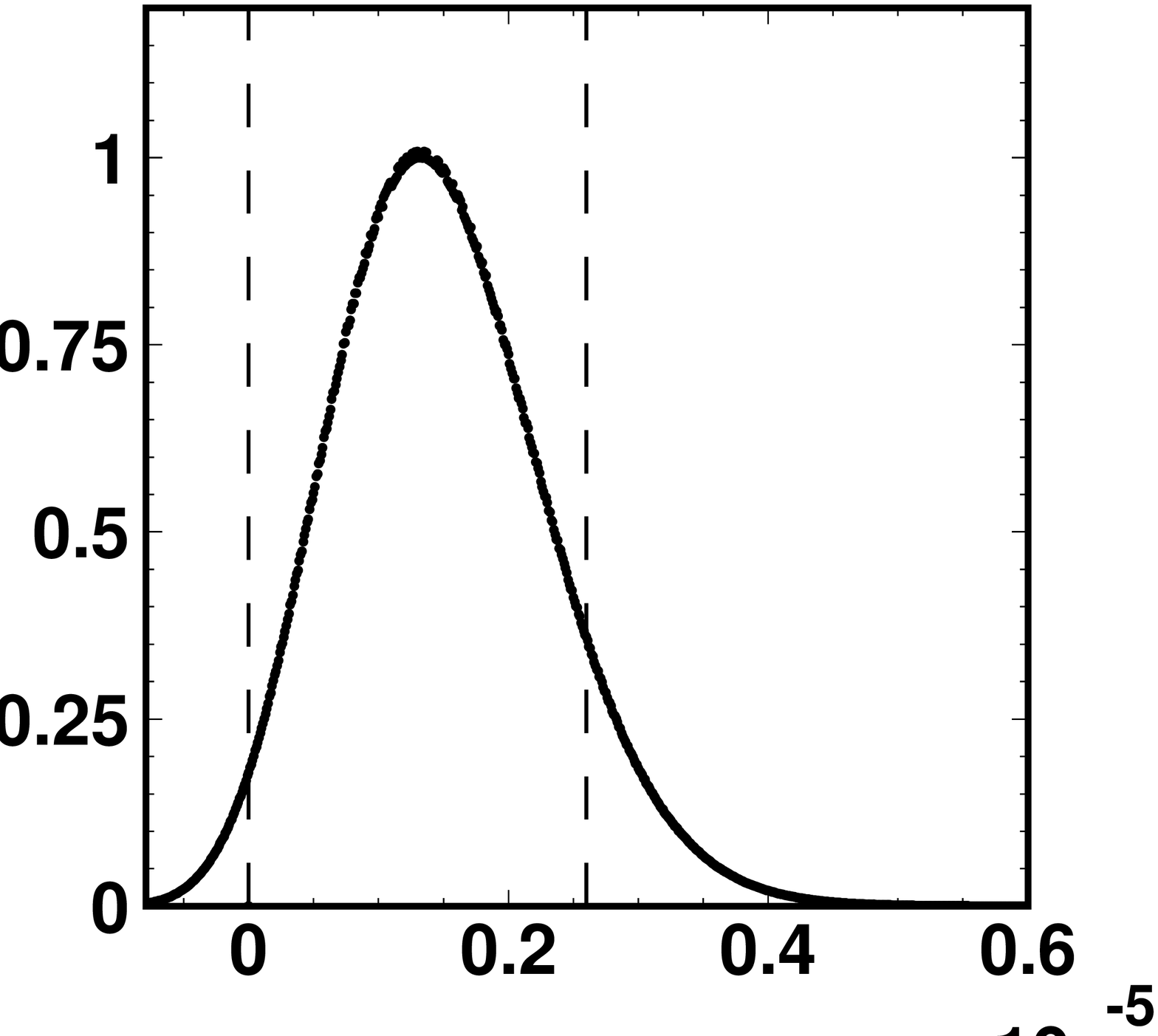}
\includegraphics[width=0.22\textwidth]{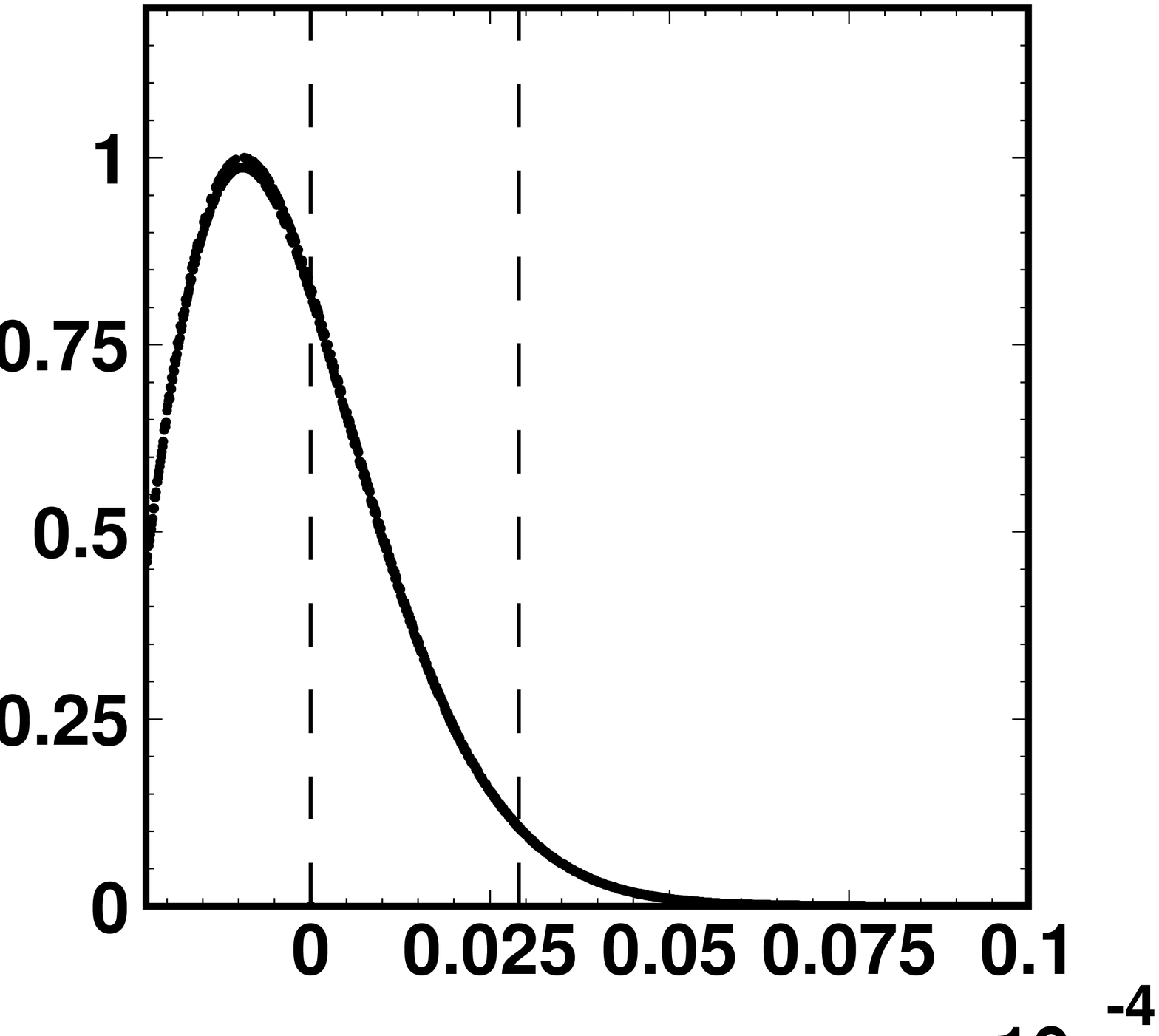} 
\includegraphics[width=0.22\textwidth]{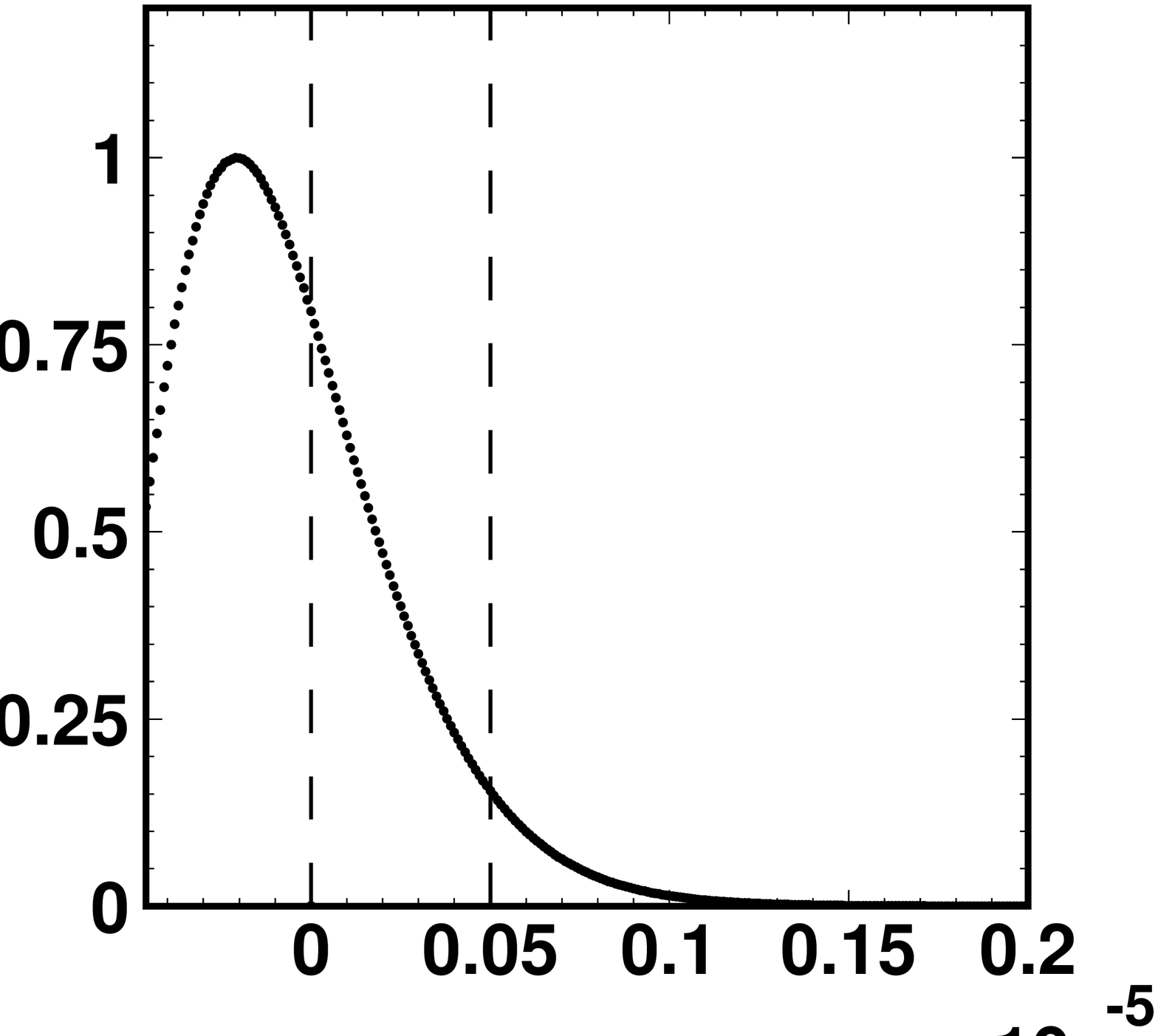}
\includegraphics[width=0.22\textwidth]{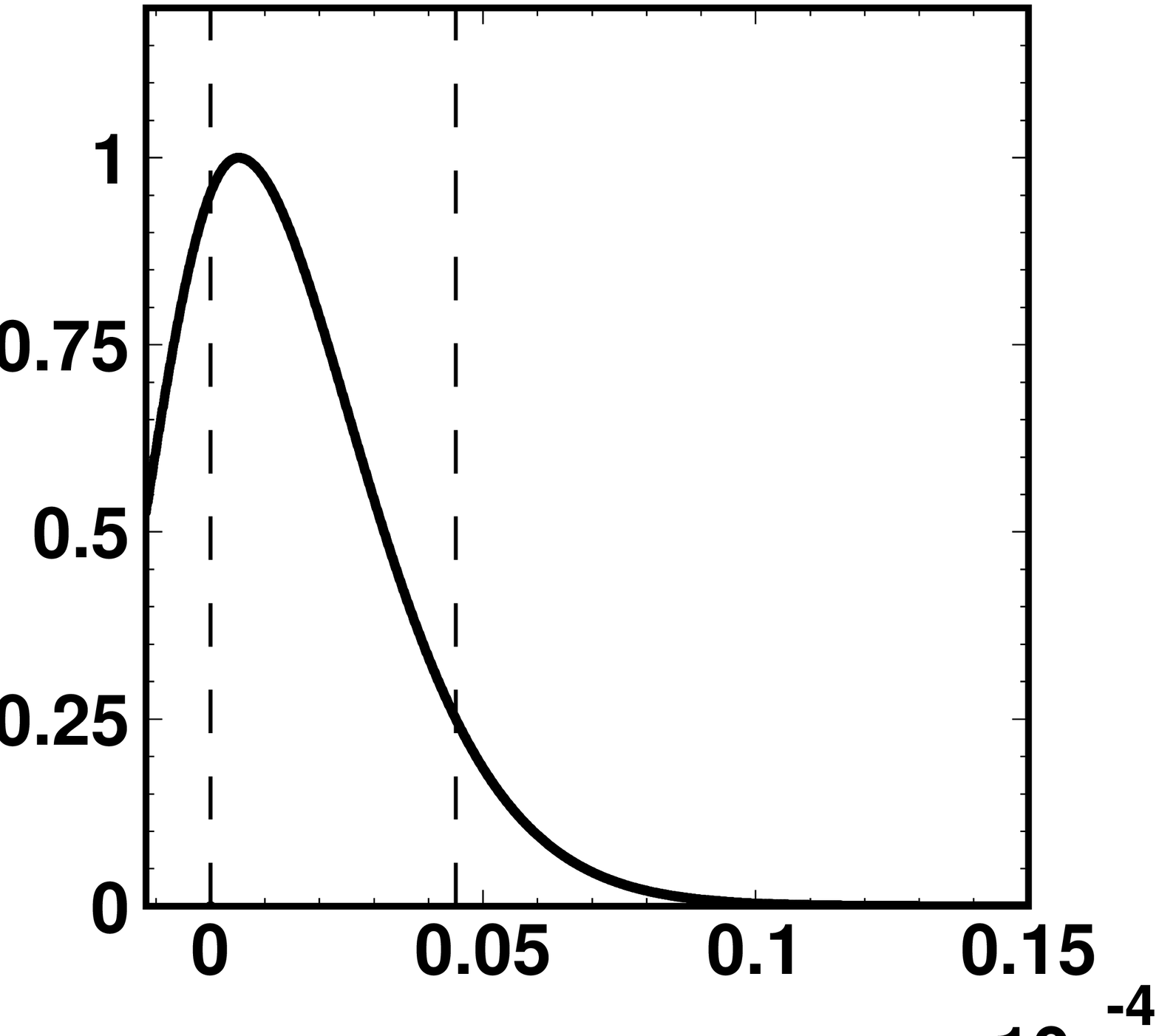}
\includegraphics[width=0.22\textwidth]{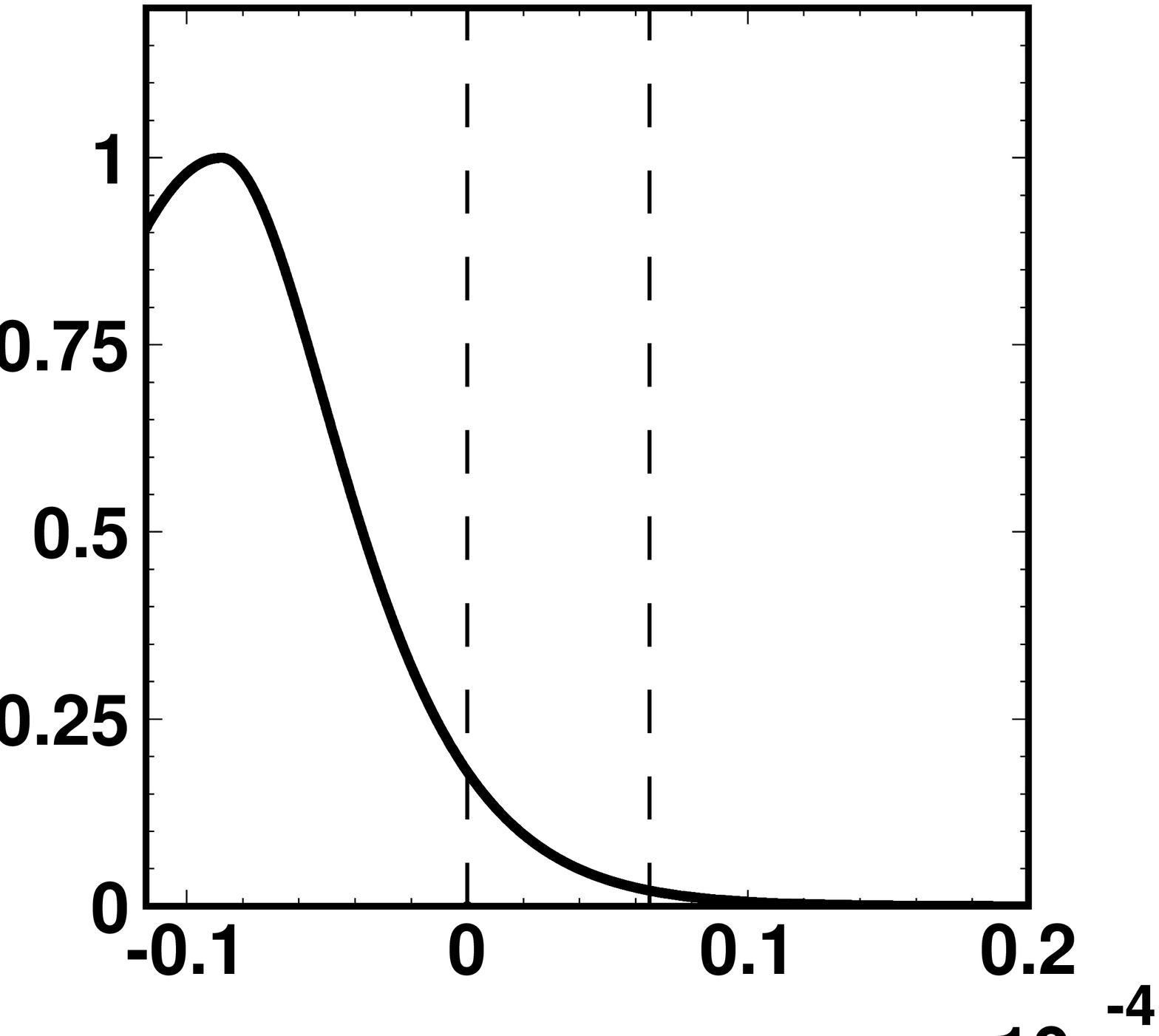}
\includegraphics[width=0.22\textwidth]{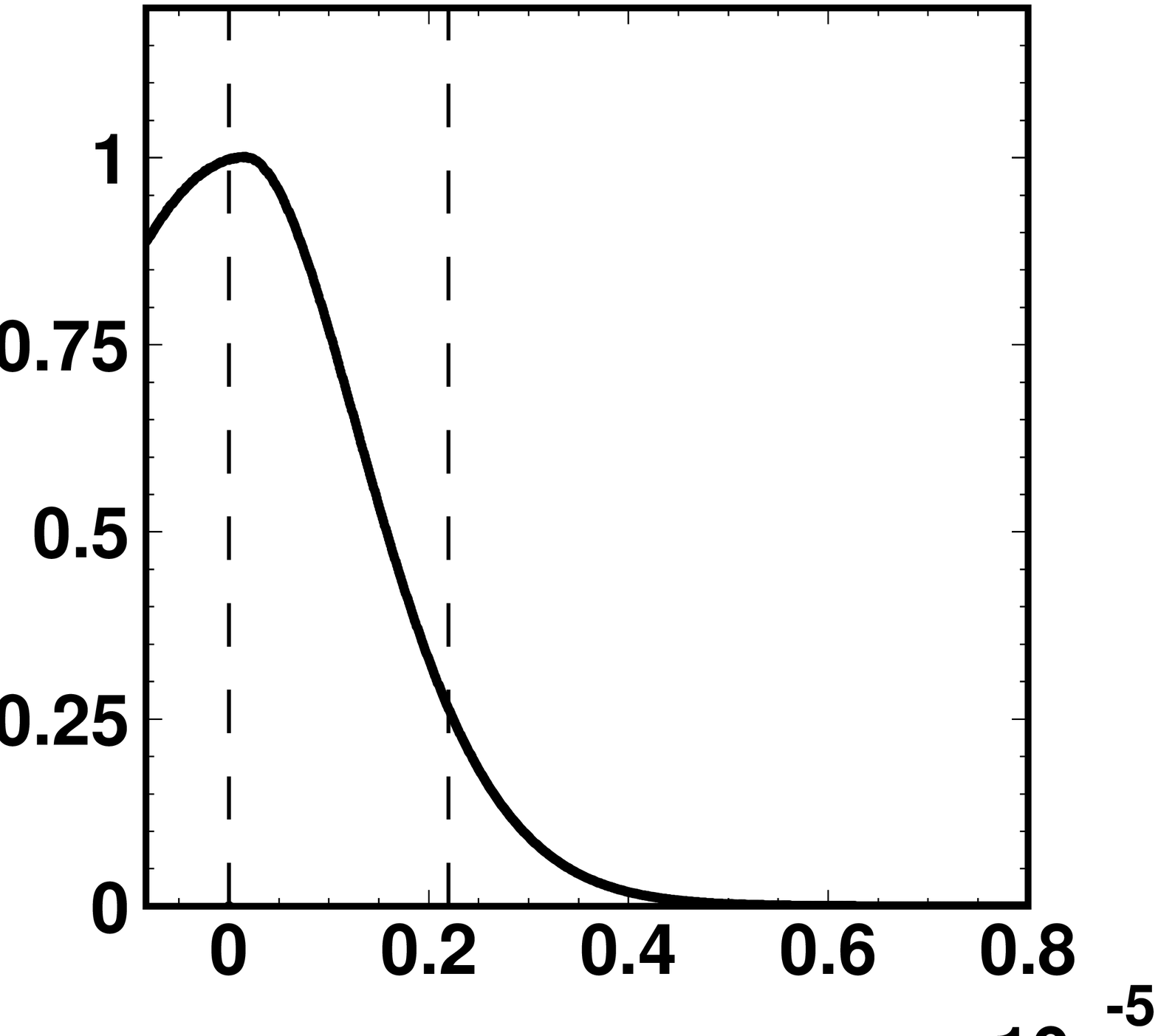}
\begin{rotate}{90}
\put(1.0,15.2){{\footnotesize{\sf\shortstack[c]{{ likelihood $L$}}}}}
\put(5.7,15.2){{\footnotesize{\sf\shortstack[c]{{ likelihood $L$}}}}}
\end{rotate}
\put(-14.1,4.1){\large{\sf\shortstack[c]{\bteprhoo}}}
\put(-10.3,4.1){\large{\sf\shortstack[c]{\bteprhop}}}
\put(-6.65,4.1){\large{\sf\shortstack[c]{\btepkstaro}}}
\put(-2.8,4.1){\large{\sf\shortstack[c]{\btepkstarp}}}
\put(-14.1,-0.6){\large{\sf\shortstack[c]{\btepphi}}}
\put(-10.3,-0.6){\large{\sf\shortstack[c]{\btepeta}}}
\put(-6.65,-0.6){\large{\sf\shortstack[c]{\btepetap}}}
\put(-2.8,-0.6){\large{\sf\shortstack[c]{\btepomega}}}
\put(-13.2,4.5){\footnotesize{\sf\shortstack[c]{\BF }}}
\put(-9.4,4.5){\footnotesize{\sf\shortstack[c]{\BF }}}
\put(-5.5,4.5){\footnotesize{\sf\shortstack[c]{\BF }}}
\put(-1.7,4.5){\footnotesize{\sf\shortstack[c]{\BF }}}
\put(-13.2,-0.2){\footnotesize{\sf\shortstack[c]{\BF }}}
\put(-9.4,-0.2){\footnotesize{\sf\shortstack[c]{\BF }}}
\put(-5.5,-0.2){\footnotesize{\sf\shortstack[c]{\BF }}}
\put(-1.7,-0.2){\footnotesize{\sf\shortstack[c]{\BF }}}
\caption{Distributions of likelihood vs. branching fraction for each decay. The 
systematic error is included as described in the text.
Two dashed lines indicate $\BF=0$ and the 90\% confidence level upper limit.}
\label{fig:lkpl}
\end{figure*}

\section{Systematics}
Systematic errors on the branching fractions 
are estimated with various high statistics data samples. 
The dominant sources are the uncertainties 
in the reconstruction efficiency of charged tracks (3--4\%), 
the uncertainties in the
reconstruction efficiencies of $\eta$ mesons, $\pi^0$'s 
and photons (3--6\%) and the $K_S^0$
reconstruction efficiency uncertainty (4\%).
Other systematic uncertainties arise from 
signal MC statistics (2\%), likelihood ratio selections (2\%),
uncertainties of the subdecay branching fractions as
given by the PDG (1.7--3.0\%), 
the number of \BB{} mesons produced (1.4\%) and 
the uncertainty from particle identification (0.5--1.3\%). 
In addition, we calculate systematic uncertainties for the fitting procedure by
varying all PDF shape parameters by $\pm 1 \sigma$.
Background normalization systematic uncertainties 
are estimated by varying the background
normalizations by 20\%-50\% while those for \de/\mb{} corrections are obtained
by varying the corrections by one standard deviation.
Since for most decays the fits yield
branching fractions close to zero, we use
absolute errors in these cases. Fractional errors are translated into absolute
values by multiplying the obtained upper limit value by the fractional error.
The combined absolute errors are decay dependent
and lie in the range (0.01 -- 4.93 )$\times 10^{-6}$. The total systematic
uncertainties are listed in Table~\ref{tab:sys}.
\begin{table*}[htb]
\caption{Total systematic uncertainties for each decay. Listed are combined
errors for fitting, efficiency related errors and the error in the number of
\BB{} events. Conservatively, we take the total systematic error 
to be the linear sum of these. All errors are in absolute values in units of
$10^{-7}$.}
\label{tab:sys}
\begin{tabular}
{@{\hspace{0.5cm}}l@{\hspace{0.5cm}}@{\hspace{0.5cm}}c@{\hspace{0.5cm}}
@{\hspace{0.5cm}}c@{\hspace{0.5cm}}@{\hspace{0.5cm}}c@{\hspace{0.5cm}}
@{\hspace{0.5cm}}c@{\hspace{0.5cm}}}
\hline\hline
Decay		& Fitting	& Efficiency	& \#\BB	& Total \\		
\hline\hline
\bteprhoo   & $^{+0.33}_{-1.76}$   & $0.07$  & $0.02$  & $^{+0.42}_{-1.85}$    \\
\bteprhop   & $^{+2.90}_{-5.53}$   & $0.32$  & $0.06$  & $^{+3.28}_{-5.91}$    \\
\btepkstaro & $^{+0.04}_{-0.03}$   & $0.16$  & $0.04$  & $^{+0.24}_{-0.23}$    \\
\btepkstarp & $^{+0.84}_{-10.10}$  & $0.21$  & $0.04$  & $^{+1.09}_{-10.35}$    \\
\btepphi    & $\pm 0.10$	   & $0.03$  & $0.01$  & $\pm 0.14$    \\
\btepeta    & $^{+2.43}_{-0.36}$   & $0.26$  & $0.05$  & $^{+2.74}_{-0.67}$    \\
\btepetap   & $^{+24.85}_{-48.94}$ & $0.29$  & $0.05$  & $^{+25.19}_{-49.28}$    \\
\btepomega  & $^{+0.58}_{-5.19}$   & $0.16$  & $0.03$  & $^{+0.77}_{-5.38}$    \\
\hline\hline
\end{tabular}
\end{table*}

\section{Upper limit calculation}
Since no decay has more than $2 \sigma$ significance~\cite{bib:sigma}, 
we calculate upper limits on the branching fractions 
by integrating the likelihood function starting at $\BF=0$ using a
Bayesian approach assuming a uniform distribution for $\BF > 0$. 
We set the upper limit when the integral
reaches 90\% of the total area under the likelihood function. 
The systematic error is
accounted for by folding the systematic error into the width of the likelihood
distribution (Eq.~\ref{eq:ns-lkhd}) when integrating the likelihood.
Thus the upper limit (UL) is calculated
with the formula:
\begin{eqnarray}
\frac{\int_{\BF=0}^{UL}{L_{\text{sys}}(N_S,N_{B_j}) d\BF}}
{\int_{\BF=0}^{1}{L_{\text{sys}}(N_S,N_{B_j}) d\BF}} = 0.9 ,
\label{eq:ULint}
\end{eqnarray}
where $L_{\text{sys}}(N_S,N_{B_j})$ is the likelihood function with its width
increased by the systematic error. The likelihood distribution is shown
in Fig.~\ref{fig:lkpl} for each decay mode.

The thus calculated upper limits are $0.5\times 10^{-6}$ for \btepphi,
$1.3\times 10^{-6}$ for \bteprhoo, 
and in the range $2.2$--$6.5\times 10^{-6}$
for other modes, as given in Tables~\ref{tab:results1} and~\ref{tab:results2}.
We note that our upper limits for \btepkstaro{} and \btepkstarp{} are below the
central values of the \babar{} measurement.

\section{Summary}
In summary, no signal was observed with more than $2 \sigma$ significance and
stringent upper limits in the range $(0.5$ -- $6.5)\times 10^{-6}$ 
for the decays \bteprho{}, \btepkstar{}, \btepphi{},
\btepetaorp and \btepomega{} have been given. All limits
except \btepeta{} are the most stringent
upper limits presently available. Our upper
limits for \btepkstar{} are below \babar's central value. 

\section{Acknowledgments}
We thank the KEKB group for the excellent operation of the
accelerator, the KEK cryogenics group for the efficient
operation of the solenoid, and the KEK computer group and
the National Institute of Informatics for valuable computing
and Super-SINET network support. We acknowledge support from
the Ministry of Education, Culture, Sports, Science, and
Technology of Japan and the Japan Society for the Promotion
of Science; the Australian Research Council and the
Australian Department of Education, Science and Training;
the National Science Foundation of China and the Knowledge
Innovation Program of the Chinese Academy of Sciences under
contract No.~10575109 and IHEP-U-503; the Department of
Science and Technology of India; 
the BK21 program of the Ministry of Education of Korea, 
the CHEP SRC program and Basic Research program 
(grant No.~R01-2005-000-10089-0) of the Korea Science and
Engineering Foundation, and the Pure Basic Research Group 
program of the Korea Research Foundation; 
the Polish State Committee for Scientific Research; 
the Ministry of Science and Technology of the Russian
Federation; the Slovenian Research Agency;  the Swiss
National Science Foundation; the National Science Council
and the Ministry of Education of Taiwan; and the U.S.\
Department of Energy.


\begin{thebibliography}{99}
\bibitem{bib:sigma}
The statistical significance of the signal
is calculated as $\sigma = \sqrt{2\ln(L_{\rm max-sys}/L_0)}$,
where $L_{0}$ and $L_{\rm max-sys}$ denote
the likelihood value at zero branching fraction and the value at one standard
deviation of the systematic error below the maximum
likelihood, respectively.
\end{thebibliography}
\end{document}